\documentclass[%
reprint,
superscriptaddress,
nofootinbib,
amsmath,amssymb,
prc,
floatfix,
]{revtex4-2}

\usepackage{graphicx}% Include figure files
\usepackage{dcolumn}% Align table columns on decimal point
\usepackage{bm}% bold math
\usepackage{color}
\usepackage{here}
\usepackage{hyperref}% add hypertext capabilities
\usepackage{nicefrac}
\usepackage{longtable}
\usepackage{lipsum}
\usepackage{ulem}
\usepackage{multirow}
\newcommand{\nuc}[2]{\hbox{$^{#1}$#2}}

\begin{document}

%\preprint{APS/123-QED}

\title{Probing proton cross-shell excitations through the two-neutron removal from \nuc{38}{Ca}}% Force line breaks with \\

%
\author{T.~Beck}
\email{beck@frib.msu.edu}
\affiliation{Facility for Rare Isotope Beams, Michigan State University, East Lansing, Michigan 48824, USA}
\author{A.~Gade}
\affiliation{Facility for Rare Isotope Beams, Michigan State University, East Lansing, Michigan 48824, USA}
\affiliation{Department of Physics and Astronomy, Michigan State University, East Lansing, Michigan 48824 USA}
\author{B.A.~Brown}
\affiliation{Facility for Rare Isotope Beams, Michigan State University, East Lansing, Michigan 48824, USA}
\affiliation{Department of Physics and Astronomy, Michigan State University, East Lansing, Michigan 48824 USA}
\author{J.A.~Tostevin}
\affiliation{Department of Physics, Faculty of Engineering and Physical Sciences, University of Surrey, Guildford, Surrey GU2 7XH, United Kingdom}
\author{D.~Weisshaar}
\affiliation{Facility for Rare Isotope Beams, Michigan State University, East Lansing, Michigan 48824, USA}
\author{D.~Bazin}
\affiliation{Facility for Rare Isotope Beams, Michigan State University, East Lansing, Michigan 48824, USA}
\affiliation{Department of Physics and Astronomy, Michigan State University, East Lansing, Michigan 48824 USA}
\author{K.W.~Brown}
\affiliation{Facility for Rare Isotope Beams, Michigan State University, East Lansing, Michigan 48824, USA}
\affiliation{Department of Chemistry, Michigan State University, East Lansing, Michigan 48824 USA}
\author{R.J.~Charity}
\affiliation{Department of Chemistry, Washington University, St. Louis, Missouri 63130 USA}
\author{P.J.~Farris}
\affiliation{Facility for Rare Isotope Beams, Michigan State University, East Lansing, Michigan 48824, USA}
\affiliation{Department of Physics and Astronomy, Michigan State University, East Lansing, Michigan 48824 USA}
\author{S.A.~Gillespie}
\affiliation{Facility for Rare Isotope Beams, Michigan State University, East Lansing, Michigan 48824, USA}
\author{A.M.~Hill}
\affiliation{Facility for Rare Isotope Beams, Michigan State University, East Lansing, Michigan 48824, USA}
\affiliation{Department of Physics and Astronomy, Michigan State University, East Lansing, Michigan 48824 USA}
\author{J.~Li}
\affiliation{Facility for Rare Isotope Beams, Michigan State University, East Lansing, Michigan 48824, USA}
\author{B.~Longfellow}
\altaffiliation[Present address: ]{Lawrence Livermore National Laboratory, Livermore, California 94550, USA}
\affiliation{Facility for Rare Isotope Beams, Michigan State University, East Lansing, Michigan 48824, USA}
\affiliation{Department of Physics and Astronomy, Michigan State University, East Lansing, Michigan 48824 USA}
\author{W.~Reviol}
\affiliation{Physics Division, Argonne National Laboratory, Argonne, Illinois 60439 USA}
\author{D.~Rhodes}
\altaffiliation[Present address: ]{TRIUMF, 4004 Wesbrook Mall, Vancouver, BC V6T 2A3, Canada}
\affiliation{Facility for Rare Isotope Beams, Michigan State University, East Lansing, Michigan 48824, USA}
\affiliation{Department of Physics and Astronomy, Michigan State University, East Lansing, Michigan 48824 USA}

\date{\today}% It is always \today, today, but any date may be explicitly specified

\begin{abstract}
Bound states of the neutron-deficient, near-dripline nucleus $^{36}$Ca were populated in two-neutron removal from the ground state of $^{38}$Ca, a direct reaction sensitive to the single-particle configurations and couplings of the removed neutrons in the projectile wave function. Final-state exclusive cross sections for the formation of \nuc{36}{Ca} and  the corresponding longitudinal momentum distributions, both determined through the combination of particle and $\gamma$-ray spectroscopy, are compared to predictions combining eikonal reaction theory and shell-model two-nucleon amplitudes from the USDB, USDC, and ZBM2 effective interactions. The final-state cross-section ratio $\sigma(2^+_1)/\sigma(0^+)$  shows particular sensitivity and is approximately reproduced only with the two-nucleon amplitudes from the ZBM2 effective interaction that includes proton cross-shell excitations into the $pf$ shell. Characterizing the proton $pf$-shell occupancy locally and schematically, an increase of the $sd$-$pf$ shell gap by $250$\,keV yields an improved description of this cross-section ratio and simultaneously enables a reproduction of the $B(E2;0^+_1\to2^+_1)$ excitation strength of $^{36}$Ca. This highlights an important aspect if a new shell-model effective interaction for the region was to be developed on the quest to model the neutron-deficient Ca isotopes and surrounding nuclei whose structure is impacted by proton cross-shell excitations.  
%\lipsum[1]
\end{abstract}

\pacs{Valid PACS appear here}% PACS, the Physics and Astronomy Classification Scheme.
%\keywords{Suggested keywords}%Use showkeys class option if keyword display desired
\maketitle

The calcium isotopic chain constitutes an attractive testing ground for nuclear models due to its diverse structure with changing neutron number, $N$. Besides the well-established magic numbers at $N=20$ and $28$, new neutron \mbox{(sub-)}shell closures have been identified for the neutron-rich $^{52}$Ca~\cite{Huc85a,Wie13a,Ros15a} and $^{54}$Ca~\cite{Ste13a,Mic18a} isotopes at $N=32$  and 34, respectively. At the neutron-deficient end of the calcium isotopic chain, a recent mass measurement of $^{35}$Ca suggests a doubly-magic character also for $^{36}$Ca ($N=16$)~\cite{Lal23a}. Due to their location at the northern boundary of the $sd$ shell, the nuclear structure of the calcium isotopes is highly susceptible to the occupation of proton $pf$-shell orbitals. However, calculations using nuclear density functional theory, which were employed for the description of nuclear charge radii, indicate only a small proton $pf$-shell occupancy in the ground state~\cite{Mil19a}, thus largely leaving the $Z=20$ shell closure intact. Studies of the nucleus $^{36}$Ca using configuration-interaction shell-model calculations~\cite{Cau05a} corroborate these results~\cite{Val18a,Lal22a} but identify an almost pure proton $2p$-$2h$ intruder configuration for the first excited $0^+_2$ state. This interesting level exhibits one of the largest known mirror-energy differences on the nuclear chart~\cite{Val18a}, which was subsequently confirmed experimentally through a combination of $(p,d)$ and $(p,t)$ transfer reactions~\cite{Lal22a}.

In contrast to the findings above, an earlier shell-model study of nuclear charge radii was interpreted to indicate a considerable weakening of the $Z=20$ and $N=20$ shell closures already in the vicinity of doubly magic $^{40}$Ca~\cite{Cau01a}. Due to the possibility of proton excitations, a weakened $Z=20$ shell closure should also result in enhanced quadrupole collectivity of the neutron-deficient calcium isotopes. Indeed, the successful description of the measured $B(E2;0^+_1\to2^+_1)$ transition strengths of $^{36,38}$Ca was shown to require the incorporation of sizable proton $pf$-shell occupancy in the ground states~\cite{Dro23a}. In contrast, the USDB Hamiltonian, which is limited to the $sd$ model space~\cite{Bro06a}, underestimates the excitation strength by about one order of magnitude, presumably due to the absence of proton excitations. Enlarging the model space to include the $pf$ shell and employing the SDPF-U-MIX effective interaction~\cite{Cau14a}, which yields a good description of the low-energy structure and mirror-energy differences of the $^{36}$Ca-$^{36}$S pair~\cite{Val18a,Lal22a}, offers only a small increase relative to the USDB results and fails to adequately account for the low-lying collectivity of $^{36,38}$Ca~\cite{Dro23a}. This quadrupole collectivity seems solely captured by the ZBM2 effective interaction in the $(1d_{3/2},2s_{1/2},1f_{7/2},2p_{3/2})$ model space~\cite{Cau01a}, which predicts proton $sd$ shell occupancies of merely $55$ and $40\%$ in the ground states of $^{36}$Ca and $^{38}$Ca, respectively~\cite{Dro23a}. In the future, the emerging evidence for the influence of proton cross-shell excitations could be tested and quantified through direct proton-removal reactions. 

Here, the direct two-neutron removal reaction from $^{38}$Ca is used to gain valuable and complementary insights into the wave-function overlap of the ground state of \nuc{38}{Ca} and the bound final states of $^{36}$Ca. The associated final-state exclusive cross sections are modeled through a combination of eikonal reaction dynamics and shell-model two-nucleon amplitudes (TNAs)~\cite{Tos04a,Tos06a}. The latter, which encode the overlap between the initial and final states,
facilitate a sensitive and unique benchmark of different model spaces and interactions when confronting the measured and theoretical cross sections. In the present case of two-neutron removal from the ground state of $^{38}$Ca to final states of $^{36}$Ca, calculated cross sections $\sigma^{\text{th}}$, employing the USDB, USDC, and ZBM2 interactions, 
are compared to experimental values from the $^{9}$Be($^{38}$Ca,$^{36}$Ca$+\gamma$)$X$ reaction to explore the adequacy of the pure $sd$-shell description (USDB and USDC) versus the need for proton cross-shell excitations across the $Z=20$ shell gap (ZBM2).

The unstable secondary $^{38}$Ca beam was produced through fragmentation of a stable $^{40}$Ca beam, accelerated by the Coupled Cyclotron Facility of the National Superconducting Cyclotron Laboratory~\cite{Gad16a} to 140\,MeV/u, on a 799-mg/cm$^2$ \nuc{9}{Be} production target and separated using a 300 mg/cm$^2$ Al wedge degrader in the A1900 fragment separator~\cite{Mor03a}. The momentum acceptance of the separator was limited to $\Delta p/p=0.25$~\%. The resulting secondary beam, which exhibited an approximately $85\%$ purity in $^{38}$Ca and a typical rate of $1.6\cdot10^5$ \nuc{38}{Ca} ions per second,
was impinged on a secondary 188\,mg/cm$^2$ thick $^9$Be target located at the target position of the S800 magnetic spectrograph~\cite{Baz03a}. The midtarget energy of the \nuc{38}{Ca} projectiles was $60.9$\,MeV/u, corresponding to a velocity of $v/c\approx0.345$.

\begin{figure}[t]
\centering
\includegraphics[trim=0 0 20 0,width=1.025\linewidth,clip]{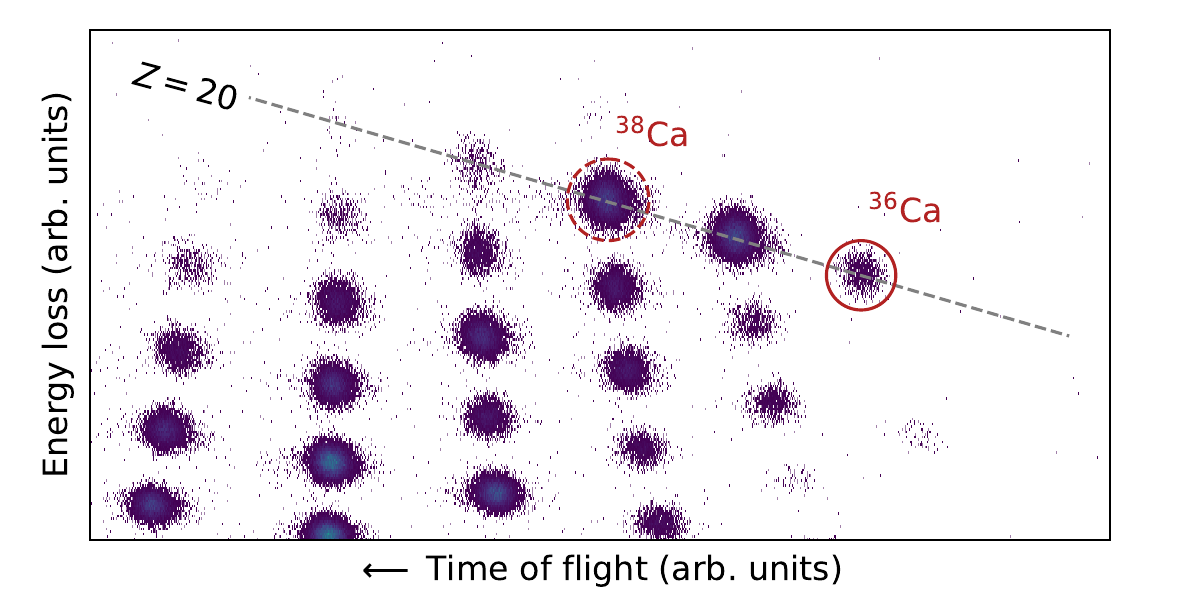}
\caption[]{Event-by-event particle identification plot of projectile-like residues produced in the $^{9}$Be($^{38}$Ca,$X+\gamma$)$Y$ reaction. The figure displays the data used for determination of absolute inclusive cross sections
and has a particle-$\gamma$ coincidence condition applied. Inelastically scattered $^{38}$Ca beam entering the focal plane 
and the $^{36}$Ca two-neutron removal residues are marked. }
\label{fig:pid}
\end{figure}

The particle identification in the entrance and exit channel was performed on an event-by-event basis using plastic timing scintillators and the S800 focal-plane detection system~\cite{Yur99a}. The incoming projectiles were identified from their velocity difference measured between two plastic scintillators located at the exit of the A1900 and the object position of the S800 analysis beam line. The outgoing reaction residues were distinguished based on their energy loss determined in the S800 ionization chamber and the time of flight measured between the object position of the analysis beam line and the plastic scintillator at the back of the S800 focal plane. As evident from the particle-identification plot shown in Fig.~\ref{fig:pid}, the $^{36}$Ca nuclei produced in two-neutron removal are well separated from the other reaction residues entering the large-acceptance focal plane of the S800 spectrograph. Results from additional reaction channels populated in this experiment are reported in Refs.~\cite{Gad20a,Gad22b,Gad22a}.

\begin{figure}[th]
\centering
\includegraphics[trim=20 0 0 0,width=1.025\linewidth,clip]{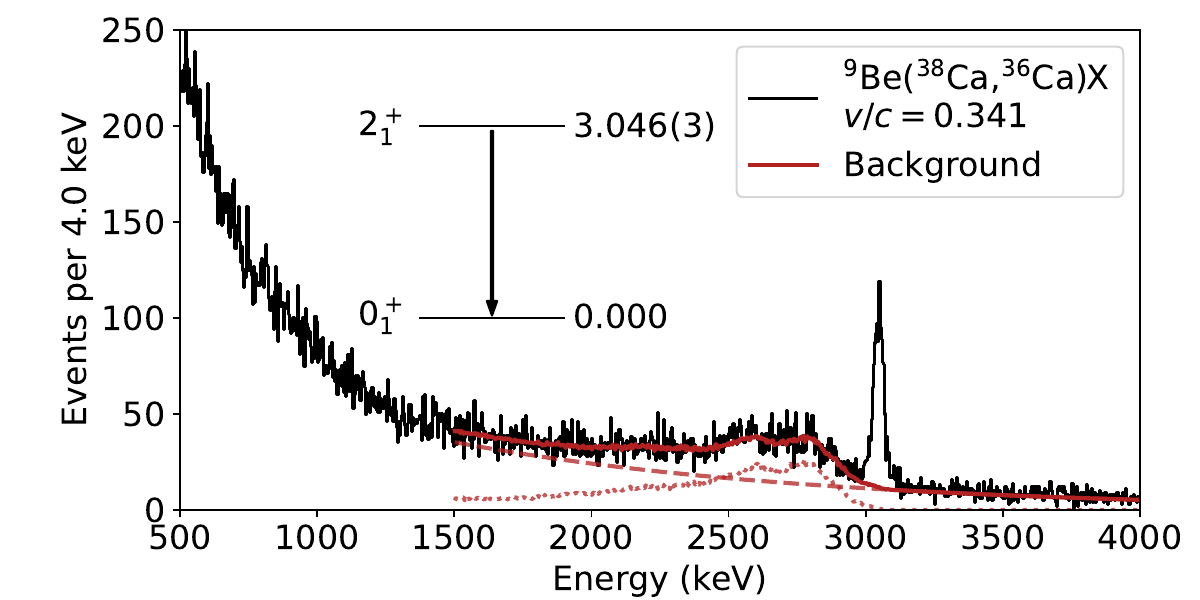}
\caption[]{Doppler-corrected ($v/c=0.341$) $\gamma$-ray spectrum detected by GRETINA
in coincidence with $^{36}$Ca residues registered in the focal plane of the S800 spectrograph.
The solid red curve represents a background model comprising an exponential component (dashed red) and 
GRETINA's detector response (dotted red) obtained from simulations employing the Geant4-based~\cite{Ago03a,All16a} code UCGretina~\cite{Ril21a}.
}
\label{fig:spec}
\end{figure}

For detection of $\gamma$ rays emitted in the decay of excited nuclear states, the reaction target was surrounded by the high-resolution $\gamma$-ray tracking array GRETINA~\cite{Pas13a,Wei17a}, consisting of twelve detector modules;
each of which contains four 36-fold segmented high-purity germanium crystals. The Doppler-corrected $\gamma$-ray spectrum of $^{36}$Ca produced in two-neutron removal from $^{38}$Ca is shown in Fig.~\ref{fig:spec}. The event-by-event Doppler reconstruction was performed with respect to the spatial coordinates of the $\gamma$-ray interaction point with the highest energy deposition as deduced from online signal decomposition. The spectrum features a prominent peak at 3046(3)\,keV, 
corresponding to the transition of the $2^+_1$ state to the ground state. Testament to the proximity of  the proton dripline, the low proton-decay threshold of $^{36}$Ca of $S_p=2600(6)$\,keV~\cite{Sur21a,Wan21a} leaves the $2^+_1$ state with a small proton decay branch in addition to the $\gamma$ emission.
In agreement with previous $\gamma$-ray spectroscopy~\cite{Doo07a,Bur12a,Dro23a}, no transitions from the other excited states reported in Ref.~\cite{Lal22a} are observed. The recently discovered $0^+_2$ state at 2.83(13)~MeV is not only the first excited state of \nuc{36}{Ca}~\cite{Lal22a} but is also located above $S_p$. Depending on the $Q$ value, it may decay via proton emission to the $3/2^+$ ground state of \nuc{35}{K} or de-excite via an $E0$ transition to the \nuc{36}{Ca} ground state. At present, the $Q$-value uncertainty exceeds 50\% and a dominance of either decay branch is possible.    

\begin{figure}[th]
\centering
\includegraphics[trim=20 0 0 0,width=1.025\linewidth,clip]{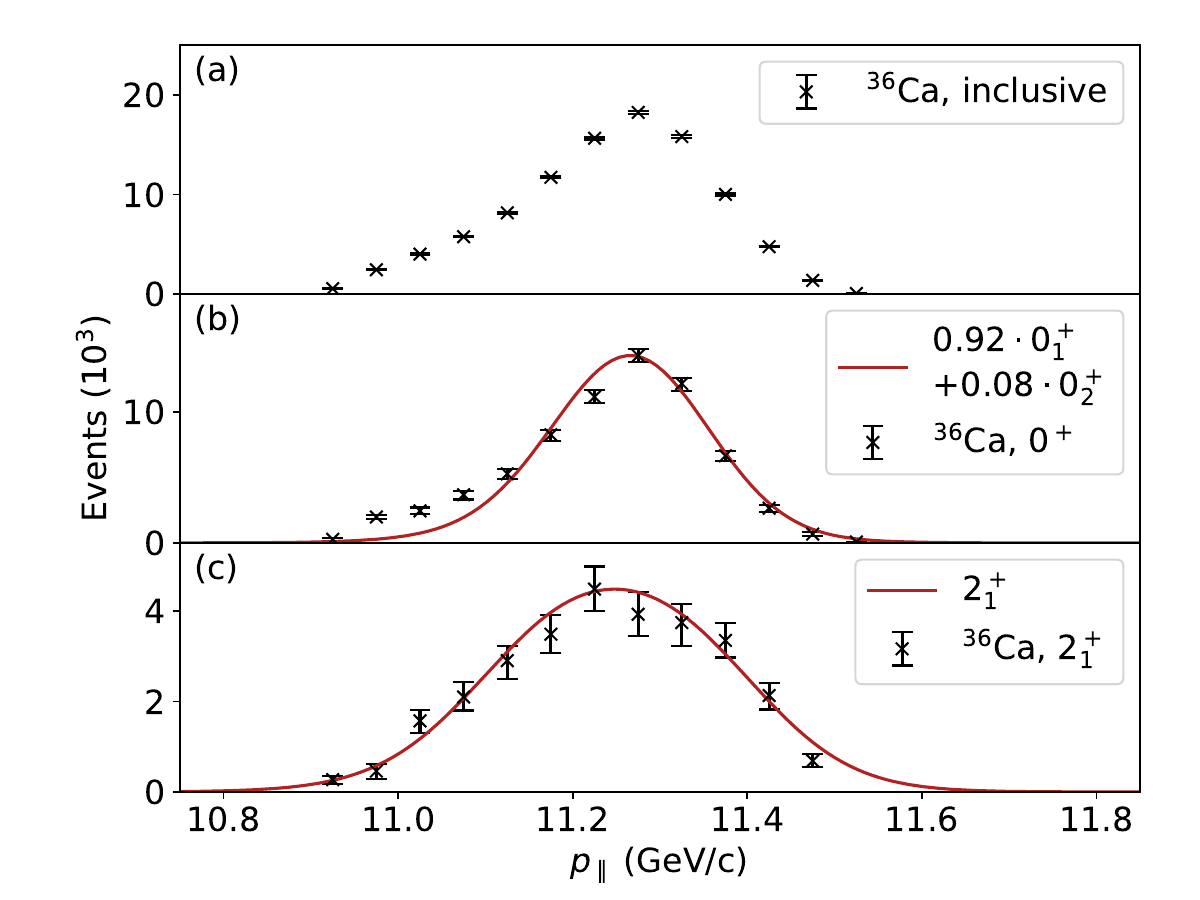}
\caption[]{Parallel momentum distributions of $^{36}$Ca after two-neutron removal from $^{38}$Ca. Panel (a) shows the inclusive $p_{\|}$ distribution which is fully within the S800 acceptance. The uncertainties are purely statistical. Panels (b) and (c) display the exclusive $p_{\|}$ distributions to the $0^+$ states and the $2^+_1$ state, respectively. They are compared to calculations using ZBM2 TNAs (red lines) which are horizontally shifted and vertically scaled for best agreement with the data. The calculation for (b) is the sum of the $0^+_1$ and $0^+_2$ contributions according to their predicted cross sections (Table~\ref{tab:xsec}). Due to the unknown proton decay branch, the $0^+_2$ contribution is an upper limit.
}
\label{fig:ppar}
\end{figure}

The S800 magnetic rigidity was optimized on $^{36}$Ca, allowing for ions with momenta of approximately $\pm0.300$\,GeV/c around central momentum $p_0=11.222$\,GeV/c to enter the spectrometer's focal plane. Fig.~\ref{fig:ppar}(a) shows the measured inclusive parallel momentum distribution of $^{36}$Ca produced in two-neutron removal from $^{38}$Ca. Employing a software gate on the observed $2^+_1\to0^+_1$ transition in the $\gamma$-ray spectrum (Fig.~\ref{fig:spec}), the partial $p_{\|}$ distribution for the  \nuc{36}{Ca} left in the $2^+_1$ final state was extracted. The parallel momentum distribution obtained after subtraction of the $2^+_1$ from the inclusive $p_{\|}$ distribution now comprises contributions from the $0^+_1$ ground state and also any contribution from the $0^+_2$ excited state if it is proton bound as the resulting $E0$ decay will escape detection. These final-state exclusive parallel momentum distributions are displayed in Fig. \ref{fig:ppar}(b) and (c).

Theoretical calculations of $p_{\|}$ distributions were obtained from eikonal reaction theory with the ZBM2 shell-model two-nucleon amplitudes as the nuclear-structure input. Since two-neutron removal from a neutron-deficient nucleus, as here, proceeds as a direct reaction~\cite{Baz03b,Yon06a,Tos04a,Tos06a}, the widths and shapes of the parallel momentum distributions have sensitivity to the total angular momentum of the removed nucleons~\cite{Sim09a,Sim09b} and also to the orbital angular momentum couplings of the two nucleons \cite{Sim10a}. Recently, the spectroscopic capabilities of $p_{\|}$ analyses were demonstrated for neutron-deficient $sd$-shell nuclei~\cite{Lon20a}. In order to facilitate a comparison to experimental $p_{\|}$ distributions, the symmetric theoretical distributions are transformed to the laboratory frame and folded with a rectangular function modeling the momentum difference due to the spatial extent of the target in the beam direction. The resulting distributions are furthermore convoluted with experimental $p_{\|}$ distributions of inelastically scattered beam particles in coincidence with $\gamma$ rays above $1000$\,keV. This last step introduces a slight asymmetry and facilitates an empirical description of the low-momentum tails frequently encountered in intermediate-energy nucleon-removal reactions~\cite{Str14a,Lon20a}. As demonstrated in  Fig.~\ref{fig:ppar}(b) and (c), the shapes of the calculated distributions obtained from this procedure agree well with the experimental ones and display the characteristic difference in widths for the different total angular momenta, $J^{\pi}$, while the low-momentum tail in the $0^+$ distribution remains slightly underestimated. 

We note that since the two-neutron removal occurs from the $J=0$ ground state of $^{38}$Ca, the total angular momentum of the removed neutrons uniquely determines the total angular momentum of the populated final state of $^{36}$Ca, confirming the previous $J^{\pi}$ assignments~\cite{Doo07a,Bur12a,Dro23a}. For the present case, the parallel momentum distributions calculated with the TNAs from the various shell-model effective interactions are essentially indistinguishable and cannot discriminate between the nuclear-structure input, unlike the cross sections discussed in the following.

The inclusive two-neutron removal cross section for the formation of $^{36}$Ca was determined from the efficiency-corrected number of $^{36}$Ca residues detected in the S800 focal plane, the number of incident $^{38}$Ca projectiles, and the areal density of the target. It amounts to $\sigma_{\text{inc}}=0.408(9)_{\text{stat}}(12)_{\text{syst}}$\,mb and is corrected for the known, small proton decay branch of the $2^+_1$ state~\cite{Dro23a}. Employing particle-$\gamma$ coincidences, the exclusive cross sections to the $0^+$ and $2^+_1$ states are obtained; the former more than twice as large as the latter (see Table \ref{tab:xsec}). Due to the low proton-separation threshold, the existence of sizable unobserved feeding can be excluded.
\begin{table*}[ht]
\caption{Comparison of experimental and calculated 
%energies of \nuc{36}{Ca} and 
two-neutron removal cross sections from $^{38}$Ca to final states of $^{36}$Ca on a $^9$Be target.
Since only the sum of cross sections to the $0^+_1$ and $0^+_2$ states, 
$\sigma(0^+)=\sigma(0^+_1)+\sigma(0^+_2)$, is experimentally accessible,
the theoretical cross sections are treated accordingly. In the case of the USDB and USDC shell-model effective interactions, the reported $0^+_2$ intruder state is outside the model space.
}
\label{tab:xsec}
\begin{ruledtabular}
\begin{tabular}{lllllll}
				&						&Experiment						&ZBM2~\cite{Cau01a}	&ZBM2gap	&USDB~\cite{Bro06a}	&USDC~\cite{Mag20a}\\\colrule
\,$\sigma$~(mb)	&$0^+_1$					&\multirow{2}{*}{$0.292(8)_{\text{stat}}(8)_{\text{syst}}$$^1$}	&0.654				&0.637		&0.609				&0.625\\
				&$0^+_2$					&								&0.058				&0.064		&				&  \\
				&$2^+_1$					&$0.116(6)_{\text{stat}}(3)_{\text{syst}}$	&0.140				&0.192		&0.824				&0.824\\
				&inclusive					&$0.408(9)_{\text{stat}}(12)_{\text{syst}}$	&0.851				&0.893		&1.433			&1.449\\\colrule
$R_s(\text{2N})$	&						&								&$0.479(10)_{\text{stat}}(14)_{\text{syst}}$	&$0.457(10)_{\text{stat}}(13)_{\text{syst}}$	&$0.285(6)_{\text{stat}}(8)_{\text{syst}}$	&$0.282(6)_{\text{stat}}(8)_{\text{syst}}$\\					
$\sigma(2^+_1)/\sigma(0^+)$		&			&$0.396(26)$						&0.197				&0.274		&1.353		&1.318\\
\end{tabular}
\end{ruledtabular}
$^1$ Combined cross section to the $0^+_1$ and $0^+_2$ states.
\end{table*}
Lacking a $\gamma$-ray detectable in GRETINA, exclusive cross sections to the $0^+_1$ and $0^+_2$ states are indistinguishable. In the following, their summed cross section is denoted $\sigma(0^+)=\sigma(0^+_1)+\sigma(0^+_2)$. In Table~\ref{tab:xsec}, the measured cross sections are compared to calculations from eikonal reaction theory incorporating shell-model TNAs from the USDB~\cite{Bro06a}, USDC~\cite{Mag20a}, and ZBM2~\cite{Cau01a} interactions. Due to the uncertain proton decay branch, the calculated two-neutron removal cross sections to the $0^+_2$ state are an upper limit.

\begin{figure}[b]
\centering
\includegraphics[trim=20 0 0 0,width=1.025\linewidth,clip]{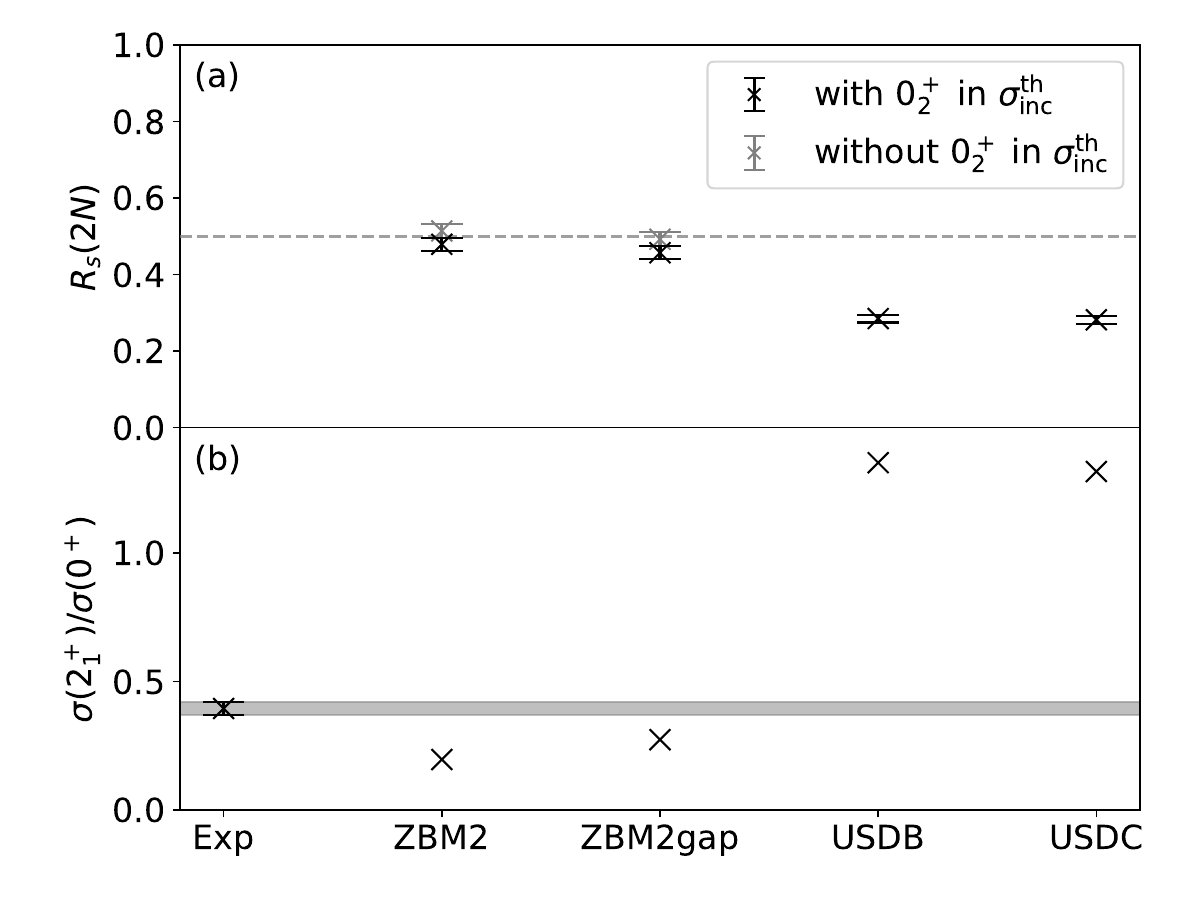}
\caption[]{Comparison of calculated and measured cross sections. (a) Measured to theoretical inclusive cross-section ratios $R_s(\text{2N})=\sigma^{\text{expt}}_{\text{inc}}/\sigma^{\text{th\vphantom{p}}}_{\text{inc}}$~\cite{Tos06a}
with (without) inclusion of the $0^+_2$ state in the calculation of the theoretical inclusive cross section
are shown in black (grey), illustrating the small contribution of the $0^+_2$ state. The expected value of $0.5$~\cite{Tos06a,Yon06a,Lon20a} is indicated by the grey dashed line. (b) Ratio of the exclusive cross sections to the $2^+_1$ and $0^+$ final states of \nuc{36}{Ca}. The $1\sigma$ region around the experimental result is indicated in gray. For this comparison, the statistical and systematic uncertainties are added in quadrature.
}
\label{fig:comp_theo}
\end{figure}

While the calculated relative cross sections of the populated final states are highly insensitive to reaction model inputs, their absolute values are reaction-model dependent. 
Differences in the absolute measured and calculated inclusive cross sections have previously been quantified in terms of their ratio 
$R_s(\text{2N})=\sigma^{\text{expt}}_{\text{inc}}/ \sigma^{ \text{th\vphantom{p}}}_{\text{inc}}$~\cite{Tos06a}. 
Earlier analyses, of several two-neutron removal cases involving nuclei embedded in the $sd$-shell, yielded values $R_s(\text{2N})\approx$ 0.5~\cite{Tos06a,Yon06a,Lon20a}. 
Although, for several two-proton removal cases, the $R_s(\text{2N})$ were significantly smaller, 
these involved systems that undergo major structural changes between the initial and final states~\cite{Gad07a,Fal10a,Cra14a,Mur19a}, not expected here.
Accounting for the experimental inseparability of $\sigma^{\text{expt}}(0^+_1)$ and $\sigma^{\text{expt}}(0^+_2)$ if the $0^+_2$ level is proton bound, 
we include cross sections to both states in the ZBM2 value of $\sigma^{\text{th}}_{\text{inc}}$ along with the $2^+_1$ state 
whose population is revealed through the $\gamma$-ray spectrum (Fig.~\ref{fig:spec}) and known proton branch~\cite{Dro23a}. 
For the $sd$-shell restricted USDB and USDC calculations the $0^+$ intruder state lies outside the model space. 
In fact, the first excited $0^+$ state calculated with the USD interactions is similar in nature to the $0^+_3$ level of ZBM2, 
almost 2~MeV above $S_p$, and would predominantly decay via proton emission. 
As evident from Fig.~\ref{fig:comp_theo}(a), calculations using the ZBM2 TNAs yield $R_s(\text{2N})$ 
consistent with earlier two-neutron removal cases while the TNAs from the USD family of effective interactions yield appreciably smaller values.

Additional guidance is obtained from the ratios of the final-state exclusive cross sections; in the present case of low proton-emission thresholds, only the cross-section ratio $\sigma(2^+_1)/\sigma(0^+)$ with an experimental value $0.396(26)$ is accessible. As shown in Fig.~\ref{fig:comp_theo}(b), the USD-type calculations limited to the $sd$ shell fail to describe the cross-section ratio. The USDB (USDC) TNAs overpredict the cross-section ratio by a factor of  3.4 (3.3) (see Table~\ref{tab:xsec} and Fig.~\ref{fig:comp_theo}). In the ZBM2 calculation, the $sd$-shell cross section is shared between the first and second $2^+$ states, the latter proton unbound, yielding the near reproduction of the observed $\sigma(2^+_1)/\sigma(0^+)$ ratio.

Figure~\ref{fig:comp_theo} immediately indicates that only the cross-section calculations with the ZBM2 TNAs get close to the data, but are not fully in agreement. 
To explore the needed amount of proton $pf$-shell occupancy schematically, we modify the $sd$-$pf$ shell gap in ZBM2.
Its size can be varied in shell-model calculations as a proxy for the magnitude of proton cross-shell excitations;
with the gap increased by $250$\,keV the fraction of proton closed-shell configuration in the ground state is raised from $55$ to $62\%$.
Using this interaction, labeled ZBM2gap in the following, the cross-section ratio $\sigma(2^+_1)/\sigma(0^+)$ of Fig.~\ref{fig:comp_theo}(b) increases from $0.20$ to $0.27$, 
reducing the deviation from experiment and hinting that potentially fewer proton cross-shell excitations than present in standard ZBM2 are needed to match the measurement.

While this is not an attempt to propose a modification of ZBM2, but rather to explore the role of proton excitations, it is instructive to study the effect on the quadrupole collectivity as well, which is expected to be very sensitive to $pf$-shell proton occupancy. Employing the ZBM2gap interaction with the same effective charges as used in Ref.~\cite{Dro23a},
$e_p=1.36$ and $e_n=0.45$, a $B(E2;0^+_1\to2^+_1)$ value of $133\,e^2$fm$^4$ for $^{36}$Ca is obtained. This value is  in better agreement with the measured transition strength of $B(E2;0^+_1\to2^+_1)=131(20)\,e^2$fm$^4$ than the calculated value of  $179\,e^2$fm$^4$ obtained with the standard ZBM2 interaction~\cite{Dro23a}. This observation is also consistent with the suggestion above that likely less proton cross-shell excitations are needed to describe the structure of \nuc{36}{Ca}. However, the significant limitations of such a local, ad-hoc modification become apparent when comparing the measured and calculated $B(E2)$ values for \nuc{38}{Ca}. The $B(E2;0^+_1\to2^+_1)= 68$$\,e^2$fm$^4$ value calculated for ZBM2gap underestimates the measured value within three sigma~\cite{Dro23a} although being in better agreement than the $sd$-shell limited calculations.             

\begin{figure}[hbt]
\centering
\includegraphics[trim=20 0 15 0,width=1.025\linewidth,clip]{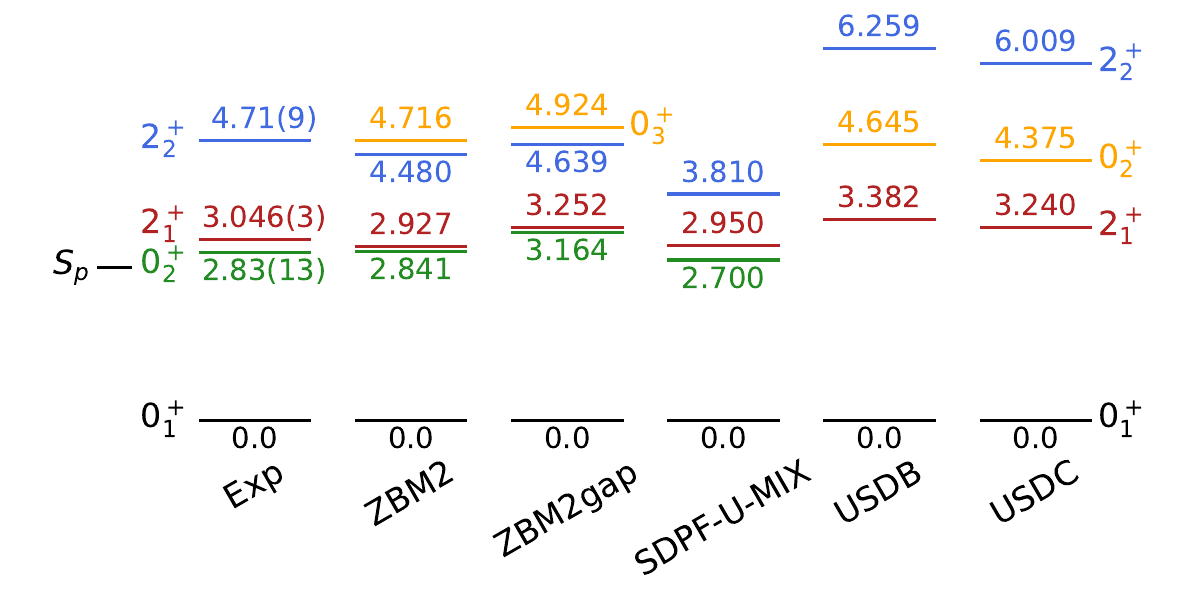}
\caption[]{Comparison of the partial experimental level scheme of $^{36}$Ca with shell-model calculations. 
The measured $0^+_2$ intruder state \cite{Lal22a} is absent from the USD-type shell-model calculations and the second excited $0^+$ state in the USD computations is closer in structure to the $0^+_3$ state of ZBM2 and ZBM2gap. The experimental excitation energies of the $0^+_2$ and $2^+_2$ states
as well as predictions from the SDPF-U-MIX interaction are taken from Ref.~\cite{Lal22a}.
Energies are given in MeV.
}
\label{fig:levels}
\end{figure}

Lastly, excitation energies of low-lying states can yield further insight into the need for proton $pf$-shell occupation. As shown in Fig.~\ref{fig:levels}, the USDB and USDC calculations do not reproduce the measured level ordering. Only the interactions incorporating proton cross-shell excitations into the $pf$ shell, namely ZBM2, ZBM2gap, and SDPF-U-MIX, correctly place the $0^+_2$ as the first excited state. In ZBM2, the $0^+_2$ level is characterized by a $42\%$ $2p$-$2h$ proton intruder component in the wave function; a weakening compared to the almost pure ($82\%$ $2p$-$2h$) intruder structure reported in Ref.~\cite{Lal22a}. A distinct difference, though, is found in the predicted energies for the $2^+_2$ state at $4.71(9)$\,MeV~\cite{Lal22a}. While calculations using the SDPF-U-MIX interaction, which identify it as an intruder state (79\% $2p$-$2h$) similar to the $0^+_2$ level, underestimate its excitation energy by approximately $0.90$\,MeV, the ZBM2 and ZBM2gap interactions enable a reproduction within $0.23$ and $0.07$\,MeV, respectively. The next $2^+$ states in ZBM2 (ZBM2gap) are $0.6$ ($0.7$)\,MeV above the ones discussed here.

Collecting the conclusions drawn above, a clear picture emerges for \nuc{36}{Ca}. Solely the ZBM2 calculations are capable of describing the low-lying level scheme and its quadrupole collectivity, and capturing the relative strengths of the final-state exclusive two-neutron removal cross sections. Furthermore, ZBM2 (ZBM2gap) provide spectroscopic factors of $0.82$ ($0.86$) and $0.42$ ($0.51$) for the one-neutron transfer from \nuc{37}{Ca} to the $0^+_1$ and $2^+_1$ states of $^{36}$Ca, respectively. It is evident that these are in better agreement with the experimental $C^2S$ values of $1.06(22)$ and $0.66(14)$ than the values $0.94$ ($0.91$) and $1.16$ ($1.01$) obtained from the USDB (SDPF-U-MIX) interactions~\cite{Lal22a}. All of this is a clear affirmation of the importance of sizeable proton $pf$-shell components in the wave functions of the neutron-deficient $^{36,38}$Ca nuclei
as proposed in Refs.~\cite{Cau01a,Dro23a}. For $^{36}$Ca, in particular the ZBM2gap interaction that slightly modifies the $pf$-shell content allows for an improved description of the experimental $B(E2;0^+_1\to2^+_1)$ and $\sigma(2^+_1)/\sigma(0^+)$ values. In this context, we note a mistake in the caption of Tab.~VI of Ref.~\cite{Dro23a}. The reported results are erroneously attributed to the standard ZBM2 interaction but were actually obtained from ZBM2gap.\footnote{For $\log_{10}(T9)<0.3$, the \nuc{35}{K}$(p,\gamma)$\nuc{36}{Ca} rate is dominated by the $2^+_1$ state for which all of the needed properties have been measured and are given in Tab.~VI of Ref.~\cite{Dro23a}. At higher temperatures the $1^+_1$ and $2^+_2$ states, which partially rely on shell-model calculations, start to contribute. Here, standard ZBM2 deviates less than $10\%$ from the ZBM2gap results.}

In summary, the $^{9}$Be($^{38}$Ca,$^{36}$Ca$+\gamma$)$X$ direct two-neutron removal reaction was used to populate bound states of $^{36}$Ca. Final-state exclusive cross sections for the formation of \nuc{36}{Ca} in the $2^+_1$ and ground state were deduced from particle-$\gamma$ coincidence spectroscopy. Eikonal reaction dynamics employing as nuclear-structure input two-nucleon amplitudes from the shell-model effective interactions USDB, USDC, and ZBM2 help clarify the importance of proton $pf$-shell occupancy: The $sd$-shell restricted interactions USDB and USDC, which describe neither the low-energy spectrum nor the low-lying quadrupole collectivity of $^{36}$Ca correctly, are also not able to reproduce the ratio of the cross sections to the two individual final states in the two-neutron removal reported here. The ZBM2 effective interaction in the  $(1d_{3/2},2s_{1/2},1f_{7/2},2p_{3/2})$ model space models the structure of the neutron-deficient calcium isotopes the best. With ZBM2 TNAs as input to the reaction model, a typical value of the measured to theoretical inclusive cross-section ratio $R_s(\text{2N})$ is obtained for the two-neutron removal from \nuc{38}{Ca} to \nuc{36}{Ca} and the ratio of the $0^+$ and $2^+_1$ exclusive cross sections is reproduced qualitatively. This agreement can be further improved by increasing the $sd$-$pf$ shell gap in ZBM2 by $250$\,keV;  this also brings  the $B(E2;0^+_1\to2^+_1)$ excitation strength of $^{36}$Ca into agreement with the experimental result. Our new studies clearly demonstrate the need for proton cross-shell excitation into the $pf$ shell already for the $0^+_1$ and $2^+_1$ states of \nuc{36}{Ca}, adding to the recent body of new data near the $sd$-shell model-space boundary. Ultimately, spectroscopic factors obtained from a proton removal reaction leading from $^{36}$Ca to (unbound) $pf$-shell final states of the very weakly-bound dripline nucleus $^{35}$K would surely quantify the $pf$-shell occupancy in the ground state of $^{36}$Ca~\cite{Dro23a}.

\begin{acknowledgments}

This work was supported by the U.S. NSF under Grants No. PHY-1565546 and No. PHY-2110365, 
by the DOE NNSA through the NSSC, under Award No. DE-NA0003180, 
and by the U.S. DOE, Office of Science, Office of Nuclear Physics, under Grants No. DE-SC0020451 \& DE-SC0023633 (MSU) and No. DE-FG02-87ER-40316 (WashU) 
and under Contract No. DE-AC02-06CH11357 (ANL).
GRETINA was funded by the DOE, Office of Science. 
Operation of the array at NSCL was supported by the DOE under Grant No. DE- SC0019034.
J. A. T. acknowledges support from the Science and Technology Facilities Council (U.K.) Grant No. ST/V001108/1.
\end{acknowledgments}

\bibliography{bib_36Ca}% Produces the bibliography via BibTeX.

\begin{thebibliography}{44}%
\makeatletter
\providecommand \@ifxundefined [1]{%
 \@ifx{#1\undefined}
}%
\providecommand \@ifnum [1]{%
 \ifnum #1\expandafter \@firstoftwo
 \else \expandafter \@secondoftwo
 \fi
}%
\providecommand \@ifx [1]{%
 \ifx #1\expandafter \@firstoftwo
 \else \expandafter \@secondoftwo
 \fi
}%
\providecommand \natexlab [1]{#1}%
\providecommand \enquote  [1]{``#1''}%
\providecommand \bibnamefont  [1]{#1}%
\providecommand \bibfnamefont [1]{#1}%
\providecommand \citenamefont [1]{#1}%
\providecommand \href@noop [0]{\@secondoftwo}%
\providecommand \href [0]{\begingroup \@sanitize@url \@href}%
\providecommand \@href[1]{\@@startlink{#1}\@@href}%
\providecommand \@@href[1]{\endgroup#1\@@endlink}%
\providecommand \@sanitize@url [0]{\catcode `\\12\catcode `\$12\catcode
  `\&12\catcode `\#12\catcode `\^12\catcode `\_12\catcode `\%12\relax}%
\providecommand \@@startlink[1]{}%
\providecommand \@@endlink[0]{}%
\providecommand \url  [0]{\begingroup\@sanitize@url \@url }%
\providecommand \@url [1]{\endgroup\@href {#1}{\urlprefix }}%
\providecommand \urlprefix  [0]{URL }%
\providecommand \Eprint [0]{\href }%
\providecommand \doibase [0]{https://doi.org/}%
\providecommand \selectlanguage [0]{\@gobble}%
\providecommand \bibinfo  [0]{\@secondoftwo}%
\providecommand \bibfield  [0]{\@secondoftwo}%
\providecommand \translation [1]{[#1]}%
\providecommand \BibitemOpen [0]{}%
\providecommand \bibitemStop [0]{}%
\providecommand \bibitemNoStop [0]{.\EOS\space}%
\providecommand \EOS [0]{\spacefactor3000\relax}%
\providecommand \BibitemShut  [1]{\csname bibitem#1\endcsname}%
\let\auto@bib@innerbib\@empty
%</preamble>
\bibitem [{\citenamefont {Huck}\ \emph {et~al.}(1985)\citenamefont {Huck},
  \citenamefont {Klotz}, \citenamefont {Knipper}, \citenamefont {Mieh\'e},
  \citenamefont {Richard-Serre}, \citenamefont {Walter}, \citenamefont {Poves},
  \citenamefont {Ravn},\ and\ \citenamefont {Marguier}}]{Huc85a}%
  \BibitemOpen
  \bibfield  {author} {\bibinfo {author} {\bibfnamefont {A.}~\bibnamefont
  {Huck}}, \bibinfo {author} {\bibfnamefont {G.}~\bibnamefont {Klotz}},
  \bibinfo {author} {\bibfnamefont {A.}~\bibnamefont {Knipper}}, \bibinfo
  {author} {\bibfnamefont {C.}~\bibnamefont {Mieh\'e}}, \bibinfo {author}
  {\bibfnamefont {C.}~\bibnamefont {Richard-Serre}}, \bibinfo {author}
  {\bibfnamefont {G.}~\bibnamefont {Walter}}, \bibinfo {author} {\bibfnamefont
  {A.}~\bibnamefont {Poves}}, \bibinfo {author} {\bibfnamefont {H.~L.}\
  \bibnamefont {Ravn}},\ and\ \bibinfo {author} {\bibfnamefont
  {G.}~\bibnamefont {Marguier}},\ }\bibfield  {title} {\bibinfo {title} {Beta
  decay of the new isotopes $^{52}\mathrm{K}$, $^{52}\mathrm{Ca}$, and
  $^{52}\mathrm{Sc}$; a test of the shell model far from stability},\ }\href
  {https://doi.org/10.1103/PhysRevC.31.2226} {\bibfield  {journal} {\bibinfo
  {journal} {Phys. Rev. C}\ }\textbf {\bibinfo {volume} {31}},\ \bibinfo
  {pages} {2226} (\bibinfo {year} {1985})}\BibitemShut {NoStop}%
\bibitem [{\citenamefont {Wienholtz}\ \emph {et~al.}(2013)\citenamefont
  {Wienholtz}, \citenamefont {Beck}, \citenamefont {Blaum}, \citenamefont
  {Borgmann}, \citenamefont {Breitenfeldt}, \citenamefont {Cakirli},
  \citenamefont {George}, \citenamefont {Herfurth}, \citenamefont {Holt},
  \citenamefont {Kowalska}, \citenamefont {Kreim}, \citenamefont {Lunney},
  \citenamefont {Manea}, \citenamefont {Men{\'e}ndez}, \citenamefont
  {Neidherr}, \citenamefont {Rosenbusch}, \citenamefont {Schweikhard},
  \citenamefont {Schwenk}, \citenamefont {Simonis}, \citenamefont {Stanja},
  \citenamefont {Wolf},\ and\ \citenamefont {Zuber}}]{Wie13a}%
  \BibitemOpen
  \bibfield  {author} {\bibinfo {author} {\bibfnamefont {F.}~\bibnamefont
  {Wienholtz}}, \bibinfo {author} {\bibfnamefont {D.}~\bibnamefont {Beck}},
  \bibinfo {author} {\bibfnamefont {K.}~\bibnamefont {Blaum}}, \bibinfo
  {author} {\bibfnamefont {C.}~\bibnamefont {Borgmann}}, \bibinfo {author}
  {\bibfnamefont {M.}~\bibnamefont {Breitenfeldt}}, \bibinfo {author}
  {\bibfnamefont {R.~B.}\ \bibnamefont {Cakirli}}, \bibinfo {author}
  {\bibfnamefont {S.}~\bibnamefont {George}}, \bibinfo {author} {\bibfnamefont
  {F.}~\bibnamefont {Herfurth}}, \bibinfo {author} {\bibfnamefont {J.~D.}\
  \bibnamefont {Holt}}, \bibinfo {author} {\bibfnamefont {M.}~\bibnamefont
  {Kowalska}}, \bibinfo {author} {\bibfnamefont {S.}~\bibnamefont {Kreim}},
  \bibinfo {author} {\bibfnamefont {D.}~\bibnamefont {Lunney}}, \bibinfo
  {author} {\bibfnamefont {V.}~\bibnamefont {Manea}}, \bibinfo {author}
  {\bibfnamefont {J.}~\bibnamefont {Men{\'e}ndez}}, \bibinfo {author}
  {\bibfnamefont {D.}~\bibnamefont {Neidherr}}, \bibinfo {author}
  {\bibfnamefont {M.}~\bibnamefont {Rosenbusch}}, \bibinfo {author}
  {\bibfnamefont {L.}~\bibnamefont {Schweikhard}}, \bibinfo {author}
  {\bibfnamefont {A.}~\bibnamefont {Schwenk}}, \bibinfo {author} {\bibfnamefont
  {J.}~\bibnamefont {Simonis}}, \bibinfo {author} {\bibfnamefont
  {J.}~\bibnamefont {Stanja}}, \bibinfo {author} {\bibfnamefont {R.~N.}\
  \bibnamefont {Wolf}},\ and\ \bibinfo {author} {\bibfnamefont
  {K.}~\bibnamefont {Zuber}},\ }\bibfield  {title} {\bibinfo {title} {Masses of
  exotic calcium isotopes pin down nuclear forces},\ }\href
  {https://doi.org/10.1038/nature12226} {\bibfield  {journal} {\bibinfo
  {journal} {Nature}\ }\textbf {\bibinfo {volume} {498}},\ \bibinfo {pages}
  {346} (\bibinfo {year} {2013})}\BibitemShut {NoStop}%
\bibitem [{\citenamefont {Rosenbusch}\ \emph {et~al.}(2015)\citenamefont
  {Rosenbusch}, \citenamefont {Ascher}, \citenamefont {Atanasov}, \citenamefont
  {Barbieri}, \citenamefont {Beck}, \citenamefont {Blaum}, \citenamefont
  {Borgmann}, \citenamefont {Breitenfeldt}, \citenamefont {Cakirli},
  \citenamefont {Cipollone}, \citenamefont {George}, \citenamefont {Herfurth},
  \citenamefont {Kowalska}, \citenamefont {Kreim}, \citenamefont {Lunney},
  \citenamefont {Manea}, \citenamefont {Navr\'atil}, \citenamefont {Neidherr},
  \citenamefont {Schweikhard}, \citenamefont {Som\`a}, \citenamefont {Stanja},
  \citenamefont {Wienholtz}, \citenamefont {Wolf},\ and\ \citenamefont
  {Zuber}}]{Ros15a}%
  \BibitemOpen
  \bibfield  {author} {\bibinfo {author} {\bibfnamefont {M.}~\bibnamefont
  {Rosenbusch}}, \bibinfo {author} {\bibfnamefont {P.}~\bibnamefont {Ascher}},
  \bibinfo {author} {\bibfnamefont {D.}~\bibnamefont {Atanasov}}, \bibinfo
  {author} {\bibfnamefont {C.}~\bibnamefont {Barbieri}}, \bibinfo {author}
  {\bibfnamefont {D.}~\bibnamefont {Beck}}, \bibinfo {author} {\bibfnamefont
  {K.}~\bibnamefont {Blaum}}, \bibinfo {author} {\bibfnamefont
  {C.}~\bibnamefont {Borgmann}}, \bibinfo {author} {\bibfnamefont
  {M.}~\bibnamefont {Breitenfeldt}}, \bibinfo {author} {\bibfnamefont {R.~B.}\
  \bibnamefont {Cakirli}}, \bibinfo {author} {\bibfnamefont {A.}~\bibnamefont
  {Cipollone}}, \bibinfo {author} {\bibfnamefont {S.}~\bibnamefont {George}},
  \bibinfo {author} {\bibfnamefont {F.}~\bibnamefont {Herfurth}}, \bibinfo
  {author} {\bibfnamefont {M.}~\bibnamefont {Kowalska}}, \bibinfo {author}
  {\bibfnamefont {S.}~\bibnamefont {Kreim}}, \bibinfo {author} {\bibfnamefont
  {D.}~\bibnamefont {Lunney}}, \bibinfo {author} {\bibfnamefont
  {V.}~\bibnamefont {Manea}}, \bibinfo {author} {\bibfnamefont
  {P.}~\bibnamefont {Navr\'atil}}, \bibinfo {author} {\bibfnamefont
  {D.}~\bibnamefont {Neidherr}}, \bibinfo {author} {\bibfnamefont
  {L.}~\bibnamefont {Schweikhard}}, \bibinfo {author} {\bibfnamefont
  {V.}~\bibnamefont {Som\`a}}, \bibinfo {author} {\bibfnamefont
  {J.}~\bibnamefont {Stanja}}, \bibinfo {author} {\bibfnamefont
  {F.}~\bibnamefont {Wienholtz}}, \bibinfo {author} {\bibfnamefont {R.~N.}\
  \bibnamefont {Wolf}},\ and\ \bibinfo {author} {\bibfnamefont
  {K.}~\bibnamefont {Zuber}},\ }\bibfield  {title} {\bibinfo {title} {Probing
  the ${N}=32$ {S}hell {C}losure below the {M}agic {P}roton {N}umber ${Z}=20$:
  {M}ass {M}easurements of the {E}xotic {I}sotopes $^{52,53}\mathrm{K}$},\
  }\href {https://doi.org/10.1103/PhysRevLett.114.202501} {\bibfield  {journal}
  {\bibinfo  {journal} {Phys. Rev. Lett.}\ }\textbf {\bibinfo {volume} {114}},\
  \bibinfo {pages} {202501} (\bibinfo {year} {2015})}\BibitemShut {NoStop}%
\bibitem [{\citenamefont {Steppenbeck}\ \emph {et~al.}(2013)\citenamefont
  {Steppenbeck}, \citenamefont {Takeuchi}, \citenamefont {Aoi}, \citenamefont
  {Doornenbal}, \citenamefont {Matsushita}, \citenamefont {Wang}, \citenamefont
  {Baba}, \citenamefont {Fukuda}, \citenamefont {Go}, \citenamefont {Honma},
  \citenamefont {Lee}, \citenamefont {Matsui}, \citenamefont {Michimasa},
  \citenamefont {Motobayashi}, \citenamefont {Nishimura}, \citenamefont
  {Otsuka}, \citenamefont {Sakurai}, \citenamefont {Shiga}, \citenamefont
  {S{\"o}derstr{\"o}m}, \citenamefont {Sumikama}, \citenamefont {Suzuki},
  \citenamefont {Taniuchi}, \citenamefont {Utsuno}, \citenamefont
  {Valiente-Dob{\'o}n},\ and\ \citenamefont {Yoneda}}]{Ste13a}%
  \BibitemOpen
  \bibfield  {author} {\bibinfo {author} {\bibfnamefont {D.}~\bibnamefont
  {Steppenbeck}}, \bibinfo {author} {\bibfnamefont {S.}~\bibnamefont
  {Takeuchi}}, \bibinfo {author} {\bibfnamefont {N.}~\bibnamefont {Aoi}},
  \bibinfo {author} {\bibfnamefont {P.}~\bibnamefont {Doornenbal}}, \bibinfo
  {author} {\bibfnamefont {M.}~\bibnamefont {Matsushita}}, \bibinfo {author}
  {\bibfnamefont {H.}~\bibnamefont {Wang}}, \bibinfo {author} {\bibfnamefont
  {H.}~\bibnamefont {Baba}}, \bibinfo {author} {\bibfnamefont {N.}~\bibnamefont
  {Fukuda}}, \bibinfo {author} {\bibfnamefont {S.}~\bibnamefont {Go}}, \bibinfo
  {author} {\bibfnamefont {M.}~\bibnamefont {Honma}}, \bibinfo {author}
  {\bibfnamefont {J.}~\bibnamefont {Lee}}, \bibinfo {author} {\bibfnamefont
  {K.}~\bibnamefont {Matsui}}, \bibinfo {author} {\bibfnamefont
  {S.}~\bibnamefont {Michimasa}}, \bibinfo {author} {\bibfnamefont
  {T.}~\bibnamefont {Motobayashi}}, \bibinfo {author} {\bibfnamefont
  {D.}~\bibnamefont {Nishimura}}, \bibinfo {author} {\bibfnamefont
  {T.}~\bibnamefont {Otsuka}}, \bibinfo {author} {\bibfnamefont
  {H.}~\bibnamefont {Sakurai}}, \bibinfo {author} {\bibfnamefont
  {Y.}~\bibnamefont {Shiga}}, \bibinfo {author} {\bibfnamefont {P.~A.}\
  \bibnamefont {S{\"o}derstr{\"o}m}}, \bibinfo {author} {\bibfnamefont
  {T.}~\bibnamefont {Sumikama}}, \bibinfo {author} {\bibfnamefont
  {H.}~\bibnamefont {Suzuki}}, \bibinfo {author} {\bibfnamefont
  {R.}~\bibnamefont {Taniuchi}}, \bibinfo {author} {\bibfnamefont
  {Y.}~\bibnamefont {Utsuno}}, \bibinfo {author} {\bibfnamefont {J.~J.}\
  \bibnamefont {Valiente-Dob{\'o}n}},\ and\ \bibinfo {author} {\bibfnamefont
  {K.}~\bibnamefont {Yoneda}},\ }\bibfield  {title} {\bibinfo {title} {Evidence
  for a new nuclear `magic number' from the level structure of
  $^{54}\mathrm{Ca}$},\ }\href {https://doi.org/10.1038/nature12522} {\bibfield
   {journal} {\bibinfo  {journal} {Nature}\ }\textbf {\bibinfo {volume}
  {502}},\ \bibinfo {pages} {207} (\bibinfo {year} {2013})}\BibitemShut
  {NoStop}%
\bibitem [{\citenamefont {Michimasa}\ \emph {et~al.}(2018)\citenamefont
  {Michimasa}, \citenamefont {Kobayashi}, \citenamefont {Kiyokawa},
  \citenamefont {Ota}, \citenamefont {Ahn}, \citenamefont {Baba}, \citenamefont
  {Berg}, \citenamefont {Dozono}, \citenamefont {Fukuda}, \citenamefont
  {Furuno}, \citenamefont {Ideguchi}, \citenamefont {Inabe}, \citenamefont
  {Kawabata}, \citenamefont {Kawase}, \citenamefont {Kisamori}, \citenamefont
  {Kobayashi}, \citenamefont {Kubo}, \citenamefont {Kubota}, \citenamefont
  {Lee}, \citenamefont {Matsushita}, \citenamefont {Miya}, \citenamefont
  {Mizukami}, \citenamefont {Nagakura}, \citenamefont {Nishimura},
  \citenamefont {Oikawa}, \citenamefont {Sakai}, \citenamefont {Shimizu},
  \citenamefont {Stolz}, \citenamefont {Suzuki}, \citenamefont {Takaki},
  \citenamefont {Takeda}, \citenamefont {Takeuchi}, \citenamefont {Tokieda},
  \citenamefont {Uesaka}, \citenamefont {Yako}, \citenamefont {Yamaguchi},
  \citenamefont {Yanagisawa}, \citenamefont {Yokoyama}, \citenamefont
  {Yoshida},\ and\ \citenamefont {Shimoura}}]{Mic18a}%
  \BibitemOpen
  \bibfield  {author} {\bibinfo {author} {\bibfnamefont {S.}~\bibnamefont
  {Michimasa}}, \bibinfo {author} {\bibfnamefont {M.}~\bibnamefont
  {Kobayashi}}, \bibinfo {author} {\bibfnamefont {Y.}~\bibnamefont {Kiyokawa}},
  \bibinfo {author} {\bibfnamefont {S.}~\bibnamefont {Ota}}, \bibinfo {author}
  {\bibfnamefont {D.~S.}\ \bibnamefont {Ahn}}, \bibinfo {author} {\bibfnamefont
  {H.}~\bibnamefont {Baba}}, \bibinfo {author} {\bibfnamefont {G.~P.~A.}\
  \bibnamefont {Berg}}, \bibinfo {author} {\bibfnamefont {M.}~\bibnamefont
  {Dozono}}, \bibinfo {author} {\bibfnamefont {N.}~\bibnamefont {Fukuda}},
  \bibinfo {author} {\bibfnamefont {T.}~\bibnamefont {Furuno}}, \bibinfo
  {author} {\bibfnamefont {E.}~\bibnamefont {Ideguchi}}, \bibinfo {author}
  {\bibfnamefont {N.}~\bibnamefont {Inabe}}, \bibinfo {author} {\bibfnamefont
  {T.}~\bibnamefont {Kawabata}}, \bibinfo {author} {\bibfnamefont
  {S.}~\bibnamefont {Kawase}}, \bibinfo {author} {\bibfnamefont
  {K.}~\bibnamefont {Kisamori}}, \bibinfo {author} {\bibfnamefont
  {K.}~\bibnamefont {Kobayashi}}, \bibinfo {author} {\bibfnamefont
  {T.}~\bibnamefont {Kubo}}, \bibinfo {author} {\bibfnamefont {Y.}~\bibnamefont
  {Kubota}}, \bibinfo {author} {\bibfnamefont {C.~S.}\ \bibnamefont {Lee}},
  \bibinfo {author} {\bibfnamefont {M.}~\bibnamefont {Matsushita}}, \bibinfo
  {author} {\bibfnamefont {H.}~\bibnamefont {Miya}}, \bibinfo {author}
  {\bibfnamefont {A.}~\bibnamefont {Mizukami}}, \bibinfo {author}
  {\bibfnamefont {H.}~\bibnamefont {Nagakura}}, \bibinfo {author}
  {\bibfnamefont {D.}~\bibnamefont {Nishimura}}, \bibinfo {author}
  {\bibfnamefont {H.}~\bibnamefont {Oikawa}}, \bibinfo {author} {\bibfnamefont
  {H.}~\bibnamefont {Sakai}}, \bibinfo {author} {\bibfnamefont
  {Y.}~\bibnamefont {Shimizu}}, \bibinfo {author} {\bibfnamefont
  {A.}~\bibnamefont {Stolz}}, \bibinfo {author} {\bibfnamefont
  {H.}~\bibnamefont {Suzuki}}, \bibinfo {author} {\bibfnamefont
  {M.}~\bibnamefont {Takaki}}, \bibinfo {author} {\bibfnamefont
  {H.}~\bibnamefont {Takeda}}, \bibinfo {author} {\bibfnamefont
  {S.}~\bibnamefont {Takeuchi}}, \bibinfo {author} {\bibfnamefont
  {H.}~\bibnamefont {Tokieda}}, \bibinfo {author} {\bibfnamefont
  {T.}~\bibnamefont {Uesaka}}, \bibinfo {author} {\bibfnamefont
  {K.}~\bibnamefont {Yako}}, \bibinfo {author} {\bibfnamefont {Y.}~\bibnamefont
  {Yamaguchi}}, \bibinfo {author} {\bibfnamefont {Y.}~\bibnamefont
  {Yanagisawa}}, \bibinfo {author} {\bibfnamefont {R.}~\bibnamefont
  {Yokoyama}}, \bibinfo {author} {\bibfnamefont {K.}~\bibnamefont {Yoshida}},\
  and\ \bibinfo {author} {\bibfnamefont {S.}~\bibnamefont {Shimoura}},\
  }\bibfield  {title} {\bibinfo {title} {Magic nature of neutrons in
  $^{54}\mathrm{Ca}$: First mass measurements of $^{55-57}\mathrm{Ca}$},\
  }\href {https://doi.org/10.1103/PhysRevLett.121.022506} {\bibfield  {journal}
  {\bibinfo  {journal} {Phys. Rev. Lett.}\ }\textbf {\bibinfo {volume} {121}},\
  \bibinfo {pages} {022506} (\bibinfo {year} {2018})}\BibitemShut {NoStop}%
\bibitem [{\citenamefont {Lalanne}\ \emph {et~al.}(2023)\citenamefont
  {Lalanne}, \citenamefont {Sorlin}, \citenamefont {Poves}, \citenamefont
  {Assi\'e}, \citenamefont {Hammache}, \citenamefont {Koyama}, \citenamefont
  {Suzuki}, \citenamefont {Flavigny}, \citenamefont {Girard-Alcindor},
  \citenamefont {Lemasson}, \citenamefont {Matta}, \citenamefont {Roger},
  \citenamefont {Beaumel}, \citenamefont {Blumenfeld}, \citenamefont {Brown},
  \citenamefont {Santos}, \citenamefont {Delaunay}, \citenamefont
  {de~S\'er\'eville}, \citenamefont {Franchoo}, \citenamefont {Gibelin},
  \citenamefont {Guillot}, \citenamefont {Kamalou}, \citenamefont {Kitamura},
  \citenamefont {Lapoux}, \citenamefont {Mauss}, \citenamefont {Morfouace},
  \citenamefont {Pancin}, \citenamefont {Saito}, \citenamefont {Stodel},\ and\
  \citenamefont {Thomas}}]{Lal23a}%
  \BibitemOpen
  \bibfield  {author} {\bibinfo {author} {\bibfnamefont {L.}~\bibnamefont
  {Lalanne}}, \bibinfo {author} {\bibfnamefont {O.}~\bibnamefont {Sorlin}},
  \bibinfo {author} {\bibfnamefont {A.}~\bibnamefont {Poves}}, \bibinfo
  {author} {\bibfnamefont {M.}~\bibnamefont {Assi\'e}}, \bibinfo {author}
  {\bibfnamefont {F.}~\bibnamefont {Hammache}}, \bibinfo {author}
  {\bibfnamefont {S.}~\bibnamefont {Koyama}}, \bibinfo {author} {\bibfnamefont
  {D.}~\bibnamefont {Suzuki}}, \bibinfo {author} {\bibfnamefont
  {F.}~\bibnamefont {Flavigny}}, \bibinfo {author} {\bibfnamefont
  {V.}~\bibnamefont {Girard-Alcindor}}, \bibinfo {author} {\bibfnamefont
  {A.}~\bibnamefont {Lemasson}}, \bibinfo {author} {\bibfnamefont
  {A.}~\bibnamefont {Matta}}, \bibinfo {author} {\bibfnamefont
  {T.}~\bibnamefont {Roger}}, \bibinfo {author} {\bibfnamefont
  {D.}~\bibnamefont {Beaumel}}, \bibinfo {author} {\bibfnamefont
  {Y.}~\bibnamefont {Blumenfeld}}, \bibinfo {author} {\bibfnamefont {B.~A.}\
  \bibnamefont {Brown}}, \bibinfo {author} {\bibfnamefont {F.~D.~O.}\
  \bibnamefont {Santos}}, \bibinfo {author} {\bibfnamefont {F.}~\bibnamefont
  {Delaunay}}, \bibinfo {author} {\bibfnamefont {N.}~\bibnamefont
  {de~S\'er\'eville}}, \bibinfo {author} {\bibfnamefont {S.}~\bibnamefont
  {Franchoo}}, \bibinfo {author} {\bibfnamefont {J.}~\bibnamefont {Gibelin}},
  \bibinfo {author} {\bibfnamefont {J.}~\bibnamefont {Guillot}}, \bibinfo
  {author} {\bibfnamefont {O.}~\bibnamefont {Kamalou}}, \bibinfo {author}
  {\bibfnamefont {N.}~\bibnamefont {Kitamura}}, \bibinfo {author}
  {\bibfnamefont {V.}~\bibnamefont {Lapoux}}, \bibinfo {author} {\bibfnamefont
  {B.}~\bibnamefont {Mauss}}, \bibinfo {author} {\bibfnamefont
  {P.}~\bibnamefont {Morfouace}}, \bibinfo {author} {\bibfnamefont
  {J.}~\bibnamefont {Pancin}}, \bibinfo {author} {\bibfnamefont {T.~Y.}\
  \bibnamefont {Saito}}, \bibinfo {author} {\bibfnamefont {C.}~\bibnamefont
  {Stodel}},\ and\ \bibinfo {author} {\bibfnamefont {J.-C.}\ \bibnamefont
  {Thomas}},\ }\bibfield  {title} {\bibinfo {title} {${N}=16$ {M}agicity
  {R}evealed at the {P}roton {D}rip {L}ine through the {S}tudy of
  $^{35}\mathrm{Ca}$},\ }\href {https://doi.org/10.1103/PhysRevLett.131.092501}
  {\bibfield  {journal} {\bibinfo  {journal} {Phys. Rev. Lett.}\ }\textbf
  {\bibinfo {volume} {131}},\ \bibinfo {pages} {092501} (\bibinfo {year}
  {2023})}\BibitemShut {NoStop}%
\bibitem [{\citenamefont {Miller}\ \emph {et~al.}(2019)\citenamefont {Miller},
  \citenamefont {Minamisono}, \citenamefont {Klose}, \citenamefont {Garand},
  \citenamefont {Kujawa}, \citenamefont {Lantis}, \citenamefont {Liu},
  \citenamefont {Maa{\ss}}, \citenamefont {Mantica}, \citenamefont
  {Nazarewicz}, \citenamefont {N{\"o}rtersh{\"a}user}, \citenamefont {Pineda},
  \citenamefont {Reinhard}, \citenamefont {Rossi}, \citenamefont {Sommer},
  \citenamefont {Sumithrarachchi}, \citenamefont {Teigelh{\"o}fer},\ and\
  \citenamefont {Watkins}}]{Mil19a}%
  \BibitemOpen
  \bibfield  {author} {\bibinfo {author} {\bibfnamefont {A.~J.}\ \bibnamefont
  {Miller}}, \bibinfo {author} {\bibfnamefont {K.}~\bibnamefont {Minamisono}},
  \bibinfo {author} {\bibfnamefont {A.}~\bibnamefont {Klose}}, \bibinfo
  {author} {\bibfnamefont {D.}~\bibnamefont {Garand}}, \bibinfo {author}
  {\bibfnamefont {C.}~\bibnamefont {Kujawa}}, \bibinfo {author} {\bibfnamefont
  {J.~D.}\ \bibnamefont {Lantis}}, \bibinfo {author} {\bibfnamefont
  {Y.}~\bibnamefont {Liu}}, \bibinfo {author} {\bibfnamefont {B.}~\bibnamefont
  {Maa{\ss}}}, \bibinfo {author} {\bibfnamefont {P.~F.}\ \bibnamefont
  {Mantica}}, \bibinfo {author} {\bibfnamefont {W.}~\bibnamefont {Nazarewicz}},
  \bibinfo {author} {\bibfnamefont {W.}~\bibnamefont {N{\"o}rtersh{\"a}user}},
  \bibinfo {author} {\bibfnamefont {S.~V.}\ \bibnamefont {Pineda}}, \bibinfo
  {author} {\bibfnamefont {P.~G.}\ \bibnamefont {Reinhard}}, \bibinfo {author}
  {\bibfnamefont {D.~M.}\ \bibnamefont {Rossi}}, \bibinfo {author}
  {\bibfnamefont {F.}~\bibnamefont {Sommer}}, \bibinfo {author} {\bibfnamefont
  {C.}~\bibnamefont {Sumithrarachchi}}, \bibinfo {author} {\bibfnamefont
  {A.}~\bibnamefont {Teigelh{\"o}fer}},\ and\ \bibinfo {author} {\bibfnamefont
  {J.}~\bibnamefont {Watkins}},\ }\bibfield  {title} {\bibinfo {title} {Proton
  superfluidity and charge radii in proton-rich calcium isotopes},\ }\href
  {https://doi.org/10.1038/s41567-019-0416-9} {\bibfield  {journal} {\bibinfo
  {journal} {Nat. Phys.}\ }\textbf {\bibinfo {volume} {15}},\ \bibinfo {pages}
  {432} (\bibinfo {year} {2019})}\BibitemShut {NoStop}%
\bibitem [{\citenamefont {Caurier}\ \emph {et~al.}(2005)\citenamefont
  {Caurier}, \citenamefont {Mart\'{\i}nez-Pinedo}, \citenamefont {Nowacki},
  \citenamefont {Poves},\ and\ \citenamefont {Zuker}}]{Cau05a}%
  \BibitemOpen
  \bibfield  {author} {\bibinfo {author} {\bibfnamefont {E.}~\bibnamefont
  {Caurier}}, \bibinfo {author} {\bibfnamefont {G.}~\bibnamefont
  {Mart\'{\i}nez-Pinedo}}, \bibinfo {author} {\bibfnamefont {F.}~\bibnamefont
  {Nowacki}}, \bibinfo {author} {\bibfnamefont {A.}~\bibnamefont {Poves}},\
  and\ \bibinfo {author} {\bibfnamefont {A.~P.}\ \bibnamefont {Zuker}},\
  }\bibfield  {title} {\bibinfo {title} {The shell model as a unified view of
  nuclear structure},\ }\href {https://doi.org/10.1103/RevModPhys.77.427}
  {\bibfield  {journal} {\bibinfo  {journal} {Rev. Mod. Phys.}\ }\textbf
  {\bibinfo {volume} {77}},\ \bibinfo {pages} {427} (\bibinfo {year}
  {2005})}\BibitemShut {NoStop}%
\bibitem [{\citenamefont {Valiente-Dob\'on}\ \emph {et~al.}(2018)\citenamefont
  {Valiente-Dob\'on}, \citenamefont {Poves}, \citenamefont {Gadea},\ and\
  \citenamefont {Fern\'andez-Dom\'{\i}nguez}}]{Val18a}%
  \BibitemOpen
  \bibfield  {author} {\bibinfo {author} {\bibfnamefont {J.~J.}\ \bibnamefont
  {Valiente-Dob\'on}}, \bibinfo {author} {\bibfnamefont {A.}~\bibnamefont
  {Poves}}, \bibinfo {author} {\bibfnamefont {A.}~\bibnamefont {Gadea}},\ and\
  \bibinfo {author} {\bibfnamefont {B.}~\bibnamefont
  {Fern\'andez-Dom\'{\i}nguez}},\ }\bibfield  {title} {\bibinfo {title} {Broken
  mirror symmetry in $^{36}\mathrm{S}$ and $^{36}\mathrm{Ca}$},\ }\href
  {https://doi.org/10.1103/PhysRevC.98.011302} {\bibfield  {journal} {\bibinfo
  {journal} {Phys. Rev. C}\ }\textbf {\bibinfo {volume} {98}},\ \bibinfo
  {pages} {011302} (\bibinfo {year} {2018})}\BibitemShut {NoStop}%
\bibitem [{\citenamefont {Lalanne}\ \emph {et~al.}(2022)\citenamefont
  {Lalanne}, \citenamefont {Sorlin}, \citenamefont {Poves}, \citenamefont
  {Assi\'e}, \citenamefont {Hammache}, \citenamefont {Koyama}, \citenamefont
  {Suzuki}, \citenamefont {Flavigny}, \citenamefont {Girard-Alcindor},
  \citenamefont {Lemasson}, \citenamefont {Matta}, \citenamefont {Roger},
  \citenamefont {Beaumel}, \citenamefont {Blumenfeld}, \citenamefont {Brown},
  \citenamefont {Santos}, \citenamefont {Delaunay}, \citenamefont
  {de~S\'er\'eville}, \citenamefont {Franchoo}, \citenamefont {Gibelin},
  \citenamefont {Guillot}, \citenamefont {Kamalou}, \citenamefont {Kitamura},
  \citenamefont {Lapoux}, \citenamefont {Mauss}, \citenamefont {Morfouace},
  \citenamefont {Niikura}, \citenamefont {Pancin}, \citenamefont {Saito},
  \citenamefont {Stodel},\ and\ \citenamefont {Thomas}}]{Lal22a}%
  \BibitemOpen
  \bibfield  {author} {\bibinfo {author} {\bibfnamefont {L.}~\bibnamefont
  {Lalanne}}, \bibinfo {author} {\bibfnamefont {O.}~\bibnamefont {Sorlin}},
  \bibinfo {author} {\bibfnamefont {A.}~\bibnamefont {Poves}}, \bibinfo
  {author} {\bibfnamefont {M.}~\bibnamefont {Assi\'e}}, \bibinfo {author}
  {\bibfnamefont {F.}~\bibnamefont {Hammache}}, \bibinfo {author}
  {\bibfnamefont {S.}~\bibnamefont {Koyama}}, \bibinfo {author} {\bibfnamefont
  {D.}~\bibnamefont {Suzuki}}, \bibinfo {author} {\bibfnamefont
  {F.}~\bibnamefont {Flavigny}}, \bibinfo {author} {\bibfnamefont
  {V.}~\bibnamefont {Girard-Alcindor}}, \bibinfo {author} {\bibfnamefont
  {A.}~\bibnamefont {Lemasson}}, \bibinfo {author} {\bibfnamefont
  {A.}~\bibnamefont {Matta}}, \bibinfo {author} {\bibfnamefont
  {T.}~\bibnamefont {Roger}}, \bibinfo {author} {\bibfnamefont
  {D.}~\bibnamefont {Beaumel}}, \bibinfo {author} {\bibfnamefont
  {Y.}~\bibnamefont {Blumenfeld}}, \bibinfo {author} {\bibfnamefont {B.~A.}\
  \bibnamefont {Brown}}, \bibinfo {author} {\bibfnamefont {F.~D.~O.}\
  \bibnamefont {Santos}}, \bibinfo {author} {\bibfnamefont {F.}~\bibnamefont
  {Delaunay}}, \bibinfo {author} {\bibfnamefont {N.}~\bibnamefont
  {de~S\'er\'eville}}, \bibinfo {author} {\bibfnamefont {S.}~\bibnamefont
  {Franchoo}}, \bibinfo {author} {\bibfnamefont {J.}~\bibnamefont {Gibelin}},
  \bibinfo {author} {\bibfnamefont {J.}~\bibnamefont {Guillot}}, \bibinfo
  {author} {\bibfnamefont {O.}~\bibnamefont {Kamalou}}, \bibinfo {author}
  {\bibfnamefont {N.}~\bibnamefont {Kitamura}}, \bibinfo {author}
  {\bibfnamefont {V.}~\bibnamefont {Lapoux}}, \bibinfo {author} {\bibfnamefont
  {B.}~\bibnamefont {Mauss}}, \bibinfo {author} {\bibfnamefont
  {P.}~\bibnamefont {Morfouace}}, \bibinfo {author} {\bibfnamefont
  {M.}~\bibnamefont {Niikura}}, \bibinfo {author} {\bibfnamefont
  {J.}~\bibnamefont {Pancin}}, \bibinfo {author} {\bibfnamefont {T.~Y.}\
  \bibnamefont {Saito}}, \bibinfo {author} {\bibfnamefont {C.}~\bibnamefont
  {Stodel}},\ and\ \bibinfo {author} {\bibfnamefont {J.-C.}\ \bibnamefont
  {Thomas}},\ }\bibfield  {title} {\bibinfo {title} {{S}tructure of
  $^{36}\mathrm{Ca}$ under the {C}oulomb {M}agnifying {G}lass},\ }\href
  {https://doi.org/10.1103/PhysRevLett.129.122501} {\bibfield  {journal}
  {\bibinfo  {journal} {Phys. Rev. Lett.}\ }\textbf {\bibinfo {volume} {129}},\
  \bibinfo {pages} {122501} (\bibinfo {year} {2022})}\BibitemShut {NoStop}%
\bibitem [{\citenamefont {Caurier}\ \emph {et~al.}(2001)\citenamefont
  {Caurier}, \citenamefont {Langanke}, \citenamefont {Mart\'inez-Pinedo},
  \citenamefont {Nowacki},\ and\ \citenamefont {Vogel}}]{Cau01a}%
  \BibitemOpen
  \bibfield  {author} {\bibinfo {author} {\bibfnamefont {E.}~\bibnamefont
  {Caurier}}, \bibinfo {author} {\bibfnamefont {K.}~\bibnamefont {Langanke}},
  \bibinfo {author} {\bibfnamefont {G.}~\bibnamefont {Mart\'inez-Pinedo}},
  \bibinfo {author} {\bibfnamefont {F.}~\bibnamefont {Nowacki}},\ and\ \bibinfo
  {author} {\bibfnamefont {P.}~\bibnamefont {Vogel}},\ }\bibfield  {title}
  {\bibinfo {title} {Shell model description of isotope shifts in calcium},\
  }\href {https://doi.org/https://doi.org/10.1016/S0370-2693(01)01246-1}
  {\bibfield  {journal} {\bibinfo  {journal} {Phys. Lett. B}\ }\textbf
  {\bibinfo {volume} {522}},\ \bibinfo {pages} {240} (\bibinfo {year}
  {2001})}\BibitemShut {NoStop}%
\bibitem [{\citenamefont {Dronchi}\ \emph {et~al.}(2023)\citenamefont
  {Dronchi}, \citenamefont {Weisshaar}, \citenamefont {Brown}, \citenamefont
  {Gade}, \citenamefont {Charity}, \citenamefont {Sobotka}, \citenamefont
  {Brown}, \citenamefont {Reviol}, \citenamefont {Bazin}, \citenamefont
  {Farris}, \citenamefont {Hill}, \citenamefont {Li}, \citenamefont
  {Longfellow}, \citenamefont {Rhodes}, \citenamefont {Paneru}, \citenamefont
  {Gillespie}, \citenamefont {Anthony}, \citenamefont {Rubino},\ and\
  \citenamefont {Biswas}}]{Dro23a}%
  \BibitemOpen
  \bibfield  {author} {\bibinfo {author} {\bibfnamefont {N.}~\bibnamefont
  {Dronchi}}, \bibinfo {author} {\bibfnamefont {D.}~\bibnamefont {Weisshaar}},
  \bibinfo {author} {\bibfnamefont {B.~A.}\ \bibnamefont {Brown}}, \bibinfo
  {author} {\bibfnamefont {A.}~\bibnamefont {Gade}}, \bibinfo {author}
  {\bibfnamefont {R.~J.}\ \bibnamefont {Charity}}, \bibinfo {author}
  {\bibfnamefont {L.~G.}\ \bibnamefont {Sobotka}}, \bibinfo {author}
  {\bibfnamefont {K.~W.}\ \bibnamefont {Brown}}, \bibinfo {author}
  {\bibfnamefont {W.}~\bibnamefont {Reviol}}, \bibinfo {author} {\bibfnamefont
  {D.}~\bibnamefont {Bazin}}, \bibinfo {author} {\bibfnamefont {P.~J.}\
  \bibnamefont {Farris}}, \bibinfo {author} {\bibfnamefont {A.~M.}\
  \bibnamefont {Hill}}, \bibinfo {author} {\bibfnamefont {J.}~\bibnamefont
  {Li}}, \bibinfo {author} {\bibfnamefont {B.}~\bibnamefont {Longfellow}},
  \bibinfo {author} {\bibfnamefont {D.}~\bibnamefont {Rhodes}}, \bibinfo
  {author} {\bibfnamefont {S.~N.}\ \bibnamefont {Paneru}}, \bibinfo {author}
  {\bibfnamefont {S.~A.}\ \bibnamefont {Gillespie}}, \bibinfo {author}
  {\bibfnamefont {A.}~\bibnamefont {Anthony}}, \bibinfo {author} {\bibfnamefont
  {E.}~\bibnamefont {Rubino}},\ and\ \bibinfo {author} {\bibfnamefont
  {S.}~\bibnamefont {Biswas}},\ }\bibfield  {title} {\bibinfo {title}
  {{M}easurement of the ${B}({E}2\ensuremath{\uparrow})$ strengths of
  $^{36}\mathrm{Ca}$ and $^{38}\mathrm{Ca}$},\ }\href
  {https://doi.org/10.1103/PhysRevC.107.034306} {\bibfield  {journal} {\bibinfo
   {journal} {Phys. Rev. C}\ }\textbf {\bibinfo {volume} {107}},\ \bibinfo
  {pages} {034306} (\bibinfo {year} {2023})}\BibitemShut {NoStop}%
\bibitem [{\citenamefont {Brown}\ and\ \citenamefont {Richter}(2006)}]{Bro06a}%
  \BibitemOpen
  \bibfield  {author} {\bibinfo {author} {\bibfnamefont {B.~A.}\ \bibnamefont
  {Brown}}\ and\ \bibinfo {author} {\bibfnamefont {W.~A.}\ \bibnamefont
  {Richter}},\ }\bibfield  {title} {\bibinfo {title} {{N}ew ``{USD}''
  {H}amiltonians for the $\mathit{sd}$ shell},\ }\href
  {https://doi.org/10.1103/PhysRevC.74.034315} {\bibfield  {journal} {\bibinfo
  {journal} {Phys. Rev. C}\ }\textbf {\bibinfo {volume} {74}},\ \bibinfo
  {pages} {034315} (\bibinfo {year} {2006})}\BibitemShut {NoStop}%
\bibitem [{\citenamefont {Caurier}\ \emph {et~al.}(2014)\citenamefont
  {Caurier}, \citenamefont {Nowacki},\ and\ \citenamefont {Poves}}]{Cau14a}%
  \BibitemOpen
  \bibfield  {author} {\bibinfo {author} {\bibfnamefont {E.}~\bibnamefont
  {Caurier}}, \bibinfo {author} {\bibfnamefont {F.}~\bibnamefont {Nowacki}},\
  and\ \bibinfo {author} {\bibfnamefont {A.}~\bibnamefont {Poves}},\ }\bibfield
   {title} {\bibinfo {title} {{M}erging of the islands of inversion at ${N}=20$
  and ${N}=28$},\ }\href {https://doi.org/10.1103/PhysRevC.90.014302}
  {\bibfield  {journal} {\bibinfo  {journal} {Phys. Rev. C}\ }\textbf {\bibinfo
  {volume} {90}},\ \bibinfo {pages} {014302} (\bibinfo {year}
  {2014})}\BibitemShut {NoStop}%
\bibitem [{\citenamefont {Tostevin}\ \emph {et~al.}(2004)\citenamefont
  {Tostevin}, \citenamefont {Podoly\'ak}, \citenamefont {Brown},\ and\
  \citenamefont {Hansen}}]{Tos04a}%
  \BibitemOpen
  \bibfield  {author} {\bibinfo {author} {\bibfnamefont {J.~A.}\ \bibnamefont
  {Tostevin}}, \bibinfo {author} {\bibfnamefont {G.}~\bibnamefont
  {Podoly\'ak}}, \bibinfo {author} {\bibfnamefont {B.~A.}\ \bibnamefont
  {Brown}},\ and\ \bibinfo {author} {\bibfnamefont {P.~G.}\ \bibnamefont
  {Hansen}},\ }\bibfield  {title} {\bibinfo {title} {Correlated two-nucleon
  stripping reactions},\ }\href {https://doi.org/10.1103/PhysRevC.70.064602}
  {\bibfield  {journal} {\bibinfo  {journal} {Phys. Rev. C}\ }\textbf {\bibinfo
  {volume} {70}},\ \bibinfo {pages} {064602} (\bibinfo {year}
  {2004})}\BibitemShut {NoStop}%
\bibitem [{\citenamefont {Tostevin}\ and\ \citenamefont
  {Brown}(2006)}]{Tos06a}%
  \BibitemOpen
  \bibfield  {author} {\bibinfo {author} {\bibfnamefont {J.~A.}\ \bibnamefont
  {Tostevin}}\ and\ \bibinfo {author} {\bibfnamefont {B.~A.}\ \bibnamefont
  {Brown}},\ }\bibfield  {title} {\bibinfo {title} {Diffraction dissociation
  contributions to two-nucleon knockout reactions and the suppression of
  shell-model strength},\ }\href {https://doi.org/10.1103/PhysRevC.74.064604}
  {\bibfield  {journal} {\bibinfo  {journal} {Phys. Rev. C}\ }\textbf {\bibinfo
  {volume} {74}},\ \bibinfo {pages} {064604} (\bibinfo {year}
  {2006})}\BibitemShut {NoStop}%
\bibitem [{\citenamefont {Gade}\ and\ \citenamefont {Sherrill}(2016)}]{Gad16a}%
  \BibitemOpen
  \bibfield  {author} {\bibinfo {author} {\bibfnamefont {A.}~\bibnamefont
  {Gade}}\ and\ \bibinfo {author} {\bibfnamefont {B.~M.}\ \bibnamefont
  {Sherrill}},\ }\bibfield  {title} {\bibinfo {title} {{NSCL} and {FRIB} at
  {M}ichigan {S}tate {U}niversity: {N}uclear science at the limits of
  stability},\ }\href {https://doi.org/10.1088/0031-8949/91/5/053003}
  {\bibfield  {journal} {\bibinfo  {journal} {Phys. Scr.}\ }\textbf {\bibinfo
  {volume} {91}},\ \bibinfo {pages} {053003} (\bibinfo {year}
  {2016})}\BibitemShut {NoStop}%
\bibitem [{\citenamefont {Morrissey}\ \emph {et~al.}(2003)\citenamefont
  {Morrissey}, \citenamefont {Sherrill}, \citenamefont {Steiner}, \citenamefont
  {Stolz},\ and\ \citenamefont {Wiedenhoever}}]{Mor03a}%
  \BibitemOpen
  \bibfield  {author} {\bibinfo {author} {\bibfnamefont {D.~J.}\ \bibnamefont
  {Morrissey}}, \bibinfo {author} {\bibfnamefont {B.~M.}\ \bibnamefont
  {Sherrill}}, \bibinfo {author} {\bibfnamefont {M.}~\bibnamefont {Steiner}},
  \bibinfo {author} {\bibfnamefont {A.}~\bibnamefont {Stolz}},\ and\ \bibinfo
  {author} {\bibfnamefont {I.}~\bibnamefont {Wiedenhoever}},\ }\bibfield
  {title} {\bibinfo {title} {Commissioning the {A}1900 projectile fragment
  separator},\ }\href
  {https://doi.org/https://doi.org/10.1016/S0168-583X(02)01895-5} {\bibfield
  {journal} {\bibinfo  {journal} {Nucl. Instrum. Methods Phys. Res. B}\
  }\textbf {\bibinfo {volume} {204}},\ \bibinfo {pages} {90} (\bibinfo {year}
  {2003})}\BibitemShut {NoStop}%
\bibitem [{\citenamefont {Bazin}\ \emph
  {et~al.}(2003{\natexlab{a}})\citenamefont {Bazin}, \citenamefont {Caggiano},
  \citenamefont {Sherrill}, \citenamefont {Yurkon},\ and\ \citenamefont
  {Zeller}}]{Baz03a}%
  \BibitemOpen
  \bibfield  {author} {\bibinfo {author} {\bibfnamefont {D.}~\bibnamefont
  {Bazin}}, \bibinfo {author} {\bibfnamefont {J.}~\bibnamefont {Caggiano}},
  \bibinfo {author} {\bibfnamefont {B.}~\bibnamefont {Sherrill}}, \bibinfo
  {author} {\bibfnamefont {J.}~\bibnamefont {Yurkon}},\ and\ \bibinfo {author}
  {\bibfnamefont {A.}~\bibnamefont {Zeller}},\ }\bibfield  {title} {\bibinfo
  {title} {The {S}800 spectrograph},\ }\href
  {https://doi.org/https://doi.org/10.1016/S0168-583X(02)02142-0} {\bibfield
  {journal} {\bibinfo  {journal} {Nucl. Instrum. Methods Phys. Res. B}\
  }\textbf {\bibinfo {volume} {204}},\ \bibinfo {pages} {629} (\bibinfo {year}
  {2003}{\natexlab{a}})}\BibitemShut {NoStop}%
\bibitem [{\citenamefont {Yurkon}\ \emph {et~al.}(1999)\citenamefont {Yurkon},
  \citenamefont {Bazin}, \citenamefont {Benenson}, \citenamefont {Morrissey},
  \citenamefont {Sherrill}, \citenamefont {Swan},\ and\ \citenamefont
  {Swanson}}]{Yur99a}%
  \BibitemOpen
  \bibfield  {author} {\bibinfo {author} {\bibfnamefont {J.}~\bibnamefont
  {Yurkon}}, \bibinfo {author} {\bibfnamefont {D.}~\bibnamefont {Bazin}},
  \bibinfo {author} {\bibfnamefont {W.}~\bibnamefont {Benenson}}, \bibinfo
  {author} {\bibfnamefont {D.~J.}\ \bibnamefont {Morrissey}}, \bibinfo {author}
  {\bibfnamefont {B.~M.}\ \bibnamefont {Sherrill}}, \bibinfo {author}
  {\bibfnamefont {D.}~\bibnamefont {Swan}},\ and\ \bibinfo {author}
  {\bibfnamefont {R.}~\bibnamefont {Swanson}},\ }\bibfield  {title} {\bibinfo
  {title} {Focal plane detector for the {S}800 high-resolution spectrometer},\
  }\href {https://doi.org/https://doi.org/10.1016/S0168-9002(98)00960-7}
  {\bibfield  {journal} {\bibinfo  {journal} {Nucl. Instrum. Methods Phys. Res.
  A}\ }\textbf {\bibinfo {volume} {422}},\ \bibinfo {pages} {291} (\bibinfo
  {year} {1999})}\BibitemShut {NoStop}%
\bibitem [{\citenamefont {Gade}\ \emph {et~al.}(2020)\citenamefont {Gade},
  \citenamefont {Weisshaar}, \citenamefont {Brown}, \citenamefont {Tostevin},
  \citenamefont {Bazin}, \citenamefont {Brown}, \citenamefont {Charity},
  \citenamefont {Farris}, \citenamefont {Hill}, \citenamefont {Li},
  \citenamefont {Longfellow}, \citenamefont {Reviol},\ and\ \citenamefont
  {Rhodes}}]{Gad20a}%
  \BibitemOpen
  \bibfield  {author} {\bibinfo {author} {\bibfnamefont {A.}~\bibnamefont
  {Gade}}, \bibinfo {author} {\bibfnamefont {D.}~\bibnamefont {Weisshaar}},
  \bibinfo {author} {\bibfnamefont {B.}~\bibnamefont {Brown}}, \bibinfo
  {author} {\bibfnamefont {J.}~\bibnamefont {Tostevin}}, \bibinfo {author}
  {\bibfnamefont {D.}~\bibnamefont {Bazin}}, \bibinfo {author} {\bibfnamefont
  {K.}~\bibnamefont {Brown}}, \bibinfo {author} {\bibfnamefont
  {R.}~\bibnamefont {Charity}}, \bibinfo {author} {\bibfnamefont
  {P.}~\bibnamefont {Farris}}, \bibinfo {author} {\bibfnamefont
  {A.}~\bibnamefont {Hill}}, \bibinfo {author} {\bibfnamefont {J.}~\bibnamefont
  {Li}}, \bibinfo {author} {\bibfnamefont {B.}~\bibnamefont {Longfellow}},
  \bibinfo {author} {\bibfnamefont {W.}~\bibnamefont {Reviol}},\ and\ \bibinfo
  {author} {\bibfnamefont {D.}~\bibnamefont {Rhodes}},\ }\bibfield  {title}
  {\bibinfo {title} {In-beam $\gamma$-ray spectroscopy at the proton dripline:
  $^{40}\mathrm{Sc}$},\ }\href
  {https://doi.org/https://doi.org/10.1016/j.physletb.2020.135637} {\bibfield
  {journal} {\bibinfo  {journal} {Phys. Lett. B}\ }\textbf {\bibinfo {volume}
  {808}},\ \bibinfo {pages} {135637} (\bibinfo {year} {2020})}\BibitemShut
  {NoStop}%
\bibitem [{\citenamefont {Gade}\ \emph
  {et~al.}(2022{\natexlab{a}})\citenamefont {Gade}, \citenamefont {Brown},
  \citenamefont {Weisshaar}, \citenamefont {Bazin}, \citenamefont {Brown},
  \citenamefont {Charity}, \citenamefont {Farris}, \citenamefont {Hill},
  \citenamefont {Li}, \citenamefont {Longfellow}, \citenamefont {Rhodes},
  \citenamefont {Reviol},\ and\ \citenamefont {Tostevin}}]{Gad22b}%
  \BibitemOpen
  \bibfield  {author} {\bibinfo {author} {\bibfnamefont {A.}~\bibnamefont
  {Gade}}, \bibinfo {author} {\bibfnamefont {B.~A.}\ \bibnamefont {Brown}},
  \bibinfo {author} {\bibfnamefont {D.}~\bibnamefont {Weisshaar}}, \bibinfo
  {author} {\bibfnamefont {D.}~\bibnamefont {Bazin}}, \bibinfo {author}
  {\bibfnamefont {K.~W.}\ \bibnamefont {Brown}}, \bibinfo {author}
  {\bibfnamefont {R.~J.}\ \bibnamefont {Charity}}, \bibinfo {author}
  {\bibfnamefont {P.}~\bibnamefont {Farris}}, \bibinfo {author} {\bibfnamefont
  {A.~M.}\ \bibnamefont {Hill}}, \bibinfo {author} {\bibfnamefont
  {J.}~\bibnamefont {Li}}, \bibinfo {author} {\bibfnamefont {B.}~\bibnamefont
  {Longfellow}}, \bibinfo {author} {\bibfnamefont {D.}~\bibnamefont {Rhodes}},
  \bibinfo {author} {\bibfnamefont {W.}~\bibnamefont {Reviol}},\ and\ \bibinfo
  {author} {\bibfnamefont {J.~A.}\ \bibnamefont {Tostevin}},\ }\bibfield
  {title} {\bibinfo {title} {Dissipative reactions with intermediate-energy
  beams: A novel approach to populate complex-structure states in rare
  isotopes},\ }\href {https://doi.org/10.1103/PhysRevLett.129.242501}
  {\bibfield  {journal} {\bibinfo  {journal} {Phys. Rev. Lett.}\ }\textbf
  {\bibinfo {volume} {129}},\ \bibinfo {pages} {242501} (\bibinfo {year}
  {2022}{\natexlab{a}})}\BibitemShut {NoStop}%
\bibitem [{\citenamefont {Gade}\ \emph
  {et~al.}(2022{\natexlab{b}})\citenamefont {Gade}, \citenamefont {Weisshaar},
  \citenamefont {Brown}, \citenamefont {Bazin}, \citenamefont {Brown},
  \citenamefont {Charity}, \citenamefont {Farris}, \citenamefont {Hill},
  \citenamefont {Li}, \citenamefont {Longfellow}, \citenamefont {Rhodes},
  \citenamefont {Reviol},\ and\ \citenamefont {Tostevin}}]{Gad22a}%
  \BibitemOpen
  \bibfield  {author} {\bibinfo {author} {\bibfnamefont {A.}~\bibnamefont
  {Gade}}, \bibinfo {author} {\bibfnamefont {D.}~\bibnamefont {Weisshaar}},
  \bibinfo {author} {\bibfnamefont {B.~A.}\ \bibnamefont {Brown}}, \bibinfo
  {author} {\bibfnamefont {D.}~\bibnamefont {Bazin}}, \bibinfo {author}
  {\bibfnamefont {K.~W.}\ \bibnamefont {Brown}}, \bibinfo {author}
  {\bibfnamefont {R.~J.}\ \bibnamefont {Charity}}, \bibinfo {author}
  {\bibfnamefont {P.}~\bibnamefont {Farris}}, \bibinfo {author} {\bibfnamefont
  {A.~M.}\ \bibnamefont {Hill}}, \bibinfo {author} {\bibfnamefont
  {J.}~\bibnamefont {Li}}, \bibinfo {author} {\bibfnamefont {B.}~\bibnamefont
  {Longfellow}}, \bibinfo {author} {\bibfnamefont {D.}~\bibnamefont {Rhodes}},
  \bibinfo {author} {\bibfnamefont {W.}~\bibnamefont {Reviol}},\ and\ \bibinfo
  {author} {\bibfnamefont {J.~A.}\ \bibnamefont {Tostevin}},\ }\bibfield
  {title} {\bibinfo {title} {Exploiting dissipative reactions to perform
  in-beam $\ensuremath{\gamma}$-ray spectroscopy of the neutron-deficient
  isotopes $^{38,39}\mathrm{Ca}$},\ }\href
  {https://doi.org/10.1103/PhysRevC.106.064303} {\bibfield  {journal} {\bibinfo
   {journal} {Phys. Rev. C}\ }\textbf {\bibinfo {volume} {106}},\ \bibinfo
  {pages} {064303} (\bibinfo {year} {2022}{\natexlab{b}})}\BibitemShut
  {NoStop}%
\bibitem [{\citenamefont {Paschalis}\ \emph {et~al.}(2013)\citenamefont
  {Paschalis}, \citenamefont {Lee}, \citenamefont {Macchiavelli}, \citenamefont
  {Campbell}, \citenamefont {Cromaz}, \citenamefont {Gros}, \citenamefont
  {Pavan}, \citenamefont {Qian}, \citenamefont {Clark}, \citenamefont
  {Crawford}, \citenamefont {Doering}, \citenamefont {Fallon}, \citenamefont
  {Lionberger}, \citenamefont {Loew}, \citenamefont {Petri}, \citenamefont
  {Stezelberger}, \citenamefont {Zimmermann}, \citenamefont {Radford},
  \citenamefont {Lagergren}, \citenamefont {Weisshaar}, \citenamefont
  {Winkler}, \citenamefont {Glasmacher}, \citenamefont {Anderson},\ and\
  \citenamefont {Beausang}}]{Pas13a}%
  \BibitemOpen
  \bibfield  {author} {\bibinfo {author} {\bibfnamefont {S.}~\bibnamefont
  {Paschalis}}, \bibinfo {author} {\bibfnamefont {I.}~\bibnamefont {Lee}},
  \bibinfo {author} {\bibfnamefont {A.}~\bibnamefont {Macchiavelli}}, \bibinfo
  {author} {\bibfnamefont {C.}~\bibnamefont {Campbell}}, \bibinfo {author}
  {\bibfnamefont {M.}~\bibnamefont {Cromaz}}, \bibinfo {author} {\bibfnamefont
  {S.}~\bibnamefont {Gros}}, \bibinfo {author} {\bibfnamefont {J.}~\bibnamefont
  {Pavan}}, \bibinfo {author} {\bibfnamefont {J.}~\bibnamefont {Qian}},
  \bibinfo {author} {\bibfnamefont {R.}~\bibnamefont {Clark}}, \bibinfo
  {author} {\bibfnamefont {H.}~\bibnamefont {Crawford}}, \bibinfo {author}
  {\bibfnamefont {D.}~\bibnamefont {Doering}}, \bibinfo {author} {\bibfnamefont
  {P.}~\bibnamefont {Fallon}}, \bibinfo {author} {\bibfnamefont
  {C.}~\bibnamefont {Lionberger}}, \bibinfo {author} {\bibfnamefont
  {T.}~\bibnamefont {Loew}}, \bibinfo {author} {\bibfnamefont {M.}~\bibnamefont
  {Petri}}, \bibinfo {author} {\bibfnamefont {T.}~\bibnamefont {Stezelberger}},
  \bibinfo {author} {\bibfnamefont {S.}~\bibnamefont {Zimmermann}}, \bibinfo
  {author} {\bibfnamefont {D.}~\bibnamefont {Radford}}, \bibinfo {author}
  {\bibfnamefont {K.}~\bibnamefont {Lagergren}}, \bibinfo {author}
  {\bibfnamefont {D.}~\bibnamefont {Weisshaar}}, \bibinfo {author}
  {\bibfnamefont {R.}~\bibnamefont {Winkler}}, \bibinfo {author} {\bibfnamefont
  {T.}~\bibnamefont {Glasmacher}}, \bibinfo {author} {\bibfnamefont
  {J.}~\bibnamefont {Anderson}},\ and\ \bibinfo {author} {\bibfnamefont
  {C.}~\bibnamefont {Beausang}},\ }\bibfield  {title} {\bibinfo {title} {The
  performance of the {G}amma-{R}ay {E}nergy {T}racking {I}n-beam {N}uclear
  {A}rray {GRETINA}},\ }\href
  {https://doi.org/https://doi.org/10.1016/j.nima.2013.01.009} {\bibfield
  {journal} {\bibinfo  {journal} {Nucl. Instrum. Methods Phys. Res. A}\
  }\textbf {\bibinfo {volume} {709}},\ \bibinfo {pages} {44} (\bibinfo {year}
  {2013})}\BibitemShut {NoStop}%
\bibitem [{\citenamefont {Weisshaar}\ \emph {et~al.}(2017)\citenamefont
  {Weisshaar}, \citenamefont {Bazin}, \citenamefont {Bender}, \citenamefont
  {Campbell}, \citenamefont {Recchia}, \citenamefont {Bader}, \citenamefont
  {Baugher}, \citenamefont {Belarge}, \citenamefont {Carpenter}, \citenamefont
  {Crawford}, \citenamefont {Cromaz}, \citenamefont {Elman}, \citenamefont
  {Fallon}, \citenamefont {Forney}, \citenamefont {Gade}, \citenamefont
  {Harker}, \citenamefont {Kobayashi}, \citenamefont {Langer}, \citenamefont
  {Lauritsen}, \citenamefont {Lee}, \citenamefont {Lemasson}, \citenamefont
  {Longfellow}, \citenamefont {Lunderberg}, \citenamefont {Macchiavelli},
  \citenamefont {Miki}, \citenamefont {Momiyama}, \citenamefont {Noji},
  \citenamefont {Radford}, \citenamefont {Scott}, \citenamefont {Sethi},
  \citenamefont {Stroberg}, \citenamefont {Sullivan}, \citenamefont {Titus},
  \citenamefont {Wiens}, \citenamefont {Williams}, \citenamefont {Wimmer},\
  and\ \citenamefont {Zhu}}]{Wei17a}%
  \BibitemOpen
  \bibfield  {author} {\bibinfo {author} {\bibfnamefont {D.}~\bibnamefont
  {Weisshaar}}, \bibinfo {author} {\bibfnamefont {D.}~\bibnamefont {Bazin}},
  \bibinfo {author} {\bibfnamefont {P.~C.}\ \bibnamefont {Bender}}, \bibinfo
  {author} {\bibfnamefont {C.~M.}\ \bibnamefont {Campbell}}, \bibinfo {author}
  {\bibfnamefont {F.}~\bibnamefont {Recchia}}, \bibinfo {author} {\bibfnamefont
  {V.}~\bibnamefont {Bader}}, \bibinfo {author} {\bibfnamefont
  {T.}~\bibnamefont {Baugher}}, \bibinfo {author} {\bibfnamefont
  {J.}~\bibnamefont {Belarge}}, \bibinfo {author} {\bibfnamefont {M.~P.}\
  \bibnamefont {Carpenter}}, \bibinfo {author} {\bibfnamefont {H.~L.}\
  \bibnamefont {Crawford}}, \bibinfo {author} {\bibfnamefont {M.}~\bibnamefont
  {Cromaz}}, \bibinfo {author} {\bibfnamefont {B.}~\bibnamefont {Elman}},
  \bibinfo {author} {\bibfnamefont {P.}~\bibnamefont {Fallon}}, \bibinfo
  {author} {\bibfnamefont {A.}~\bibnamefont {Forney}}, \bibinfo {author}
  {\bibfnamefont {A.}~\bibnamefont {Gade}}, \bibinfo {author} {\bibfnamefont
  {J.}~\bibnamefont {Harker}}, \bibinfo {author} {\bibfnamefont
  {N.}~\bibnamefont {Kobayashi}}, \bibinfo {author} {\bibfnamefont
  {C.}~\bibnamefont {Langer}}, \bibinfo {author} {\bibfnamefont
  {T.}~\bibnamefont {Lauritsen}}, \bibinfo {author} {\bibfnamefont {I.~Y.}\
  \bibnamefont {Lee}}, \bibinfo {author} {\bibfnamefont {A.}~\bibnamefont
  {Lemasson}}, \bibinfo {author} {\bibfnamefont {B.}~\bibnamefont
  {Longfellow}}, \bibinfo {author} {\bibfnamefont {E.}~\bibnamefont
  {Lunderberg}}, \bibinfo {author} {\bibfnamefont {A.~O.}\ \bibnamefont
  {Macchiavelli}}, \bibinfo {author} {\bibfnamefont {K.}~\bibnamefont {Miki}},
  \bibinfo {author} {\bibfnamefont {S.}~\bibnamefont {Momiyama}}, \bibinfo
  {author} {\bibfnamefont {S.}~\bibnamefont {Noji}}, \bibinfo {author}
  {\bibfnamefont {D.~C.}\ \bibnamefont {Radford}}, \bibinfo {author}
  {\bibfnamefont {M.}~\bibnamefont {Scott}}, \bibinfo {author} {\bibfnamefont
  {J.}~\bibnamefont {Sethi}}, \bibinfo {author} {\bibfnamefont {S.~R.}\
  \bibnamefont {Stroberg}}, \bibinfo {author} {\bibfnamefont {C.}~\bibnamefont
  {Sullivan}}, \bibinfo {author} {\bibfnamefont {R.}~\bibnamefont {Titus}},
  \bibinfo {author} {\bibfnamefont {A.}~\bibnamefont {Wiens}}, \bibinfo
  {author} {\bibfnamefont {S.}~\bibnamefont {Williams}}, \bibinfo {author}
  {\bibfnamefont {K.}~\bibnamefont {Wimmer}},\ and\ \bibinfo {author}
  {\bibfnamefont {S.}~\bibnamefont {Zhu}},\ }\bibfield  {title} {\bibinfo
  {title} {The performance of the $\gamma$-ray tracking array {GRETINA} for
  $\gamma$-ray spectroscopy with fast beams of rare isotopes},\ }\href
  {https://doi.org/https://doi.org/10.1016/j.nima.2016.12.001} {\bibfield
  {journal} {\bibinfo  {journal} {Nucl. Instrum. Methods Phys. Res. A}\
  }\textbf {\bibinfo {volume} {847}},\ \bibinfo {pages} {187} (\bibinfo {year}
  {2017})}\BibitemShut {NoStop}%
\bibitem [{\citenamefont {Surbrook}\ \emph {et~al.}(2021)\citenamefont
  {Surbrook}, \citenamefont {Bollen}, \citenamefont {Brodeur}, \citenamefont
  {Hamaker}, \citenamefont {P\'erez-Loureiro}, \citenamefont {Puentes},
  \citenamefont {Nicoloff}, \citenamefont {Redshaw}, \citenamefont {Ringle},
  \citenamefont {Schwarz}, \citenamefont {Sumithrarachchi}, \citenamefont
  {Sun}, \citenamefont {Valverde}, \citenamefont {Villari}, \citenamefont
  {Wrede},\ and\ \citenamefont {Yandow}}]{Sur21a}%
  \BibitemOpen
  \bibfield  {author} {\bibinfo {author} {\bibfnamefont {J.}~\bibnamefont
  {Surbrook}}, \bibinfo {author} {\bibfnamefont {G.}~\bibnamefont {Bollen}},
  \bibinfo {author} {\bibfnamefont {M.}~\bibnamefont {Brodeur}}, \bibinfo
  {author} {\bibfnamefont {A.}~\bibnamefont {Hamaker}}, \bibinfo {author}
  {\bibfnamefont {D.}~\bibnamefont {P\'erez-Loureiro}}, \bibinfo {author}
  {\bibfnamefont {D.}~\bibnamefont {Puentes}}, \bibinfo {author} {\bibfnamefont
  {C.}~\bibnamefont {Nicoloff}}, \bibinfo {author} {\bibfnamefont
  {M.}~\bibnamefont {Redshaw}}, \bibinfo {author} {\bibfnamefont
  {R.}~\bibnamefont {Ringle}}, \bibinfo {author} {\bibfnamefont
  {S.}~\bibnamefont {Schwarz}}, \bibinfo {author} {\bibfnamefont {C.~S.}\
  \bibnamefont {Sumithrarachchi}}, \bibinfo {author} {\bibfnamefont {L.~J.}\
  \bibnamefont {Sun}}, \bibinfo {author} {\bibfnamefont {A.~A.}\ \bibnamefont
  {Valverde}}, \bibinfo {author} {\bibfnamefont {A.~C.~C.}\ \bibnamefont
  {Villari}}, \bibinfo {author} {\bibfnamefont {C.}~\bibnamefont {Wrede}},\
  and\ \bibinfo {author} {\bibfnamefont {I.~T.}\ \bibnamefont {Yandow}},\
  }\bibfield  {title} {\bibinfo {title} {First {P}enning trap mass measurement
  of $^{36}\mathrm{Ca}$},\ }\href {https://doi.org/10.1103/PhysRevC.103.014323}
  {\bibfield  {journal} {\bibinfo  {journal} {Phys. Rev. C}\ }\textbf {\bibinfo
  {volume} {103}},\ \bibinfo {pages} {014323} (\bibinfo {year}
  {2021})}\BibitemShut {NoStop}%
\bibitem [{\citenamefont {Wang}\ \emph {et~al.}(2021)\citenamefont {Wang},
  \citenamefont {Huang}, \citenamefont {Kondev}, \citenamefont {Audi},\ and\
  \citenamefont {Naimi}}]{Wan21a}%
  \BibitemOpen
  \bibfield  {author} {\bibinfo {author} {\bibfnamefont {M.}~\bibnamefont
  {Wang}}, \bibinfo {author} {\bibfnamefont {W.~J.}\ \bibnamefont {Huang}},
  \bibinfo {author} {\bibfnamefont {F.~G.}\ \bibnamefont {Kondev}}, \bibinfo
  {author} {\bibfnamefont {G.}~\bibnamefont {Audi}},\ and\ \bibinfo {author}
  {\bibfnamefont {S.}~\bibnamefont {Naimi}},\ }\bibfield  {title} {\bibinfo
  {title} {The {AME} 2020 atomic mass evaluation (ii). {T}ables, graphs and
  references*},\ }\href {https://doi.org/10.1088/1674-1137/abddaf} {\bibfield
  {journal} {\bibinfo  {journal} {Chin. Phys. C}\ }\textbf {\bibinfo {volume}
  {45}},\ \bibinfo {pages} {030003} (\bibinfo {year} {2021})}\BibitemShut
  {NoStop}%
\bibitem [{\citenamefont {Doornenbal}\ \emph {et~al.}(2007)\citenamefont
  {Doornenbal}, \citenamefont {Reiter}, \citenamefont {Grawe}, \citenamefont
  {Otsuka}, \citenamefont {Al-Khatib}, \citenamefont {Banu}, \citenamefont
  {Beck}, \citenamefont {Becker}, \citenamefont {Bednarczyk}, \citenamefont
  {Benzoni}, \citenamefont {Bracco}, \citenamefont {B{\"u}rger}, \citenamefont
  {Caceres}, \citenamefont {Camera}, \citenamefont {Chmel}, \citenamefont
  {Crespi}, \citenamefont {Geissel}, \citenamefont {Gerl}, \citenamefont
  {G{\'o}rska}, \citenamefont {Gr{\c e}bosz}, \citenamefont {H{\"u}bel},
  \citenamefont {Kavatsyuk}, \citenamefont {Kavatsyuk}, \citenamefont
  {Kmiecik}, \citenamefont {Kojouharov}, \citenamefont {Kurz}, \citenamefont
  {Lozeva}, \citenamefont {Maj}, \citenamefont {Mandal}, \citenamefont
  {Meczynski}, \citenamefont {Million}, \citenamefont {Podoly{\'a}k},
  \citenamefont {Richard}, \citenamefont {Saito}, \citenamefont {Saito},
  \citenamefont {Schaffner}, \citenamefont {Seidlitz}, \citenamefont
  {Striepling}, \citenamefont {Utsuno}, \citenamefont {Walker}, \citenamefont
  {Warr}, \citenamefont {Weick}, \citenamefont {Wieland}, \citenamefont
  {Winkler},\ and\ \citenamefont {Wollersheim}}]{Doo07a}%
  \BibitemOpen
  \bibfield  {author} {\bibinfo {author} {\bibfnamefont {P.}~\bibnamefont
  {Doornenbal}}, \bibinfo {author} {\bibfnamefont {P.}~\bibnamefont {Reiter}},
  \bibinfo {author} {\bibfnamefont {H.}~\bibnamefont {Grawe}}, \bibinfo
  {author} {\bibfnamefont {T.}~\bibnamefont {Otsuka}}, \bibinfo {author}
  {\bibfnamefont {A.}~\bibnamefont {Al-Khatib}}, \bibinfo {author}
  {\bibfnamefont {A.}~\bibnamefont {Banu}}, \bibinfo {author} {\bibfnamefont
  {T.}~\bibnamefont {Beck}}, \bibinfo {author} {\bibfnamefont {F.}~\bibnamefont
  {Becker}}, \bibinfo {author} {\bibfnamefont {P.}~\bibnamefont {Bednarczyk}},
  \bibinfo {author} {\bibfnamefont {G.}~\bibnamefont {Benzoni}}, \bibinfo
  {author} {\bibfnamefont {A.}~\bibnamefont {Bracco}}, \bibinfo {author}
  {\bibfnamefont {A.}~\bibnamefont {B{\"u}rger}}, \bibinfo {author}
  {\bibfnamefont {L.}~\bibnamefont {Caceres}}, \bibinfo {author} {\bibfnamefont
  {F.}~\bibnamefont {Camera}}, \bibinfo {author} {\bibfnamefont
  {S.}~\bibnamefont {Chmel}}, \bibinfo {author} {\bibfnamefont
  {F.}~\bibnamefont {Crespi}}, \bibinfo {author} {\bibfnamefont
  {H.}~\bibnamefont {Geissel}}, \bibinfo {author} {\bibfnamefont
  {J.}~\bibnamefont {Gerl}}, \bibinfo {author} {\bibfnamefont {M.}~\bibnamefont
  {G{\'o}rska}}, \bibinfo {author} {\bibfnamefont {J.}~\bibnamefont {Gr{\c
  e}bosz}}, \bibinfo {author} {\bibfnamefont {H.}~\bibnamefont {H{\"u}bel}},
  \bibinfo {author} {\bibfnamefont {M.}~\bibnamefont {Kavatsyuk}}, \bibinfo
  {author} {\bibfnamefont {O.}~\bibnamefont {Kavatsyuk}}, \bibinfo {author}
  {\bibfnamefont {M.}~\bibnamefont {Kmiecik}}, \bibinfo {author} {\bibfnamefont
  {I.}~\bibnamefont {Kojouharov}}, \bibinfo {author} {\bibfnamefont
  {N.}~\bibnamefont {Kurz}}, \bibinfo {author} {\bibfnamefont {R.}~\bibnamefont
  {Lozeva}}, \bibinfo {author} {\bibfnamefont {A.}~\bibnamefont {Maj}},
  \bibinfo {author} {\bibfnamefont {S.}~\bibnamefont {Mandal}}, \bibinfo
  {author} {\bibfnamefont {W.}~\bibnamefont {Meczynski}}, \bibinfo {author}
  {\bibfnamefont {B.}~\bibnamefont {Million}}, \bibinfo {author} {\bibfnamefont
  {Z.}~\bibnamefont {Podoly{\'a}k}}, \bibinfo {author} {\bibfnamefont
  {A.}~\bibnamefont {Richard}}, \bibinfo {author} {\bibfnamefont
  {N.}~\bibnamefont {Saito}}, \bibinfo {author} {\bibfnamefont
  {T.}~\bibnamefont {Saito}}, \bibinfo {author} {\bibfnamefont
  {H.}~\bibnamefont {Schaffner}}, \bibinfo {author} {\bibfnamefont
  {M.}~\bibnamefont {Seidlitz}}, \bibinfo {author} {\bibfnamefont
  {T.}~\bibnamefont {Striepling}}, \bibinfo {author} {\bibfnamefont
  {Y.}~\bibnamefont {Utsuno}}, \bibinfo {author} {\bibfnamefont
  {J.}~\bibnamefont {Walker}}, \bibinfo {author} {\bibfnamefont
  {N.}~\bibnamefont {Warr}}, \bibinfo {author} {\bibfnamefont {H.}~\bibnamefont
  {Weick}}, \bibinfo {author} {\bibfnamefont {O.}~\bibnamefont {Wieland}},
  \bibinfo {author} {\bibfnamefont {M.}~\bibnamefont {Winkler}},\ and\ \bibinfo
  {author} {\bibfnamefont {H.}~\bibnamefont {Wollersheim}},\ }\bibfield
  {title} {\bibinfo {title} {The ${T}=2$ mirrors $^{36}\mathrm{Ca}$ and
  $^{36}\mathrm{S}$: {A} test for isospin symmetry of shell gaps at the
  driplines},\ }\href
  {https://doi.org/https://doi.org/10.1016/j.physletb.2007.02.001} {\bibfield
  {journal} {\bibinfo  {journal} {Phys. Lett. B}\ }\textbf {\bibinfo {volume}
  {647}},\ \bibinfo {pages} {237} (\bibinfo {year} {2007})}\BibitemShut
  {NoStop}%
\bibitem [{\citenamefont {B\"urger}\ \emph {et~al.}(2012)\citenamefont
  {B\"urger}, \citenamefont {Azaiez}, \citenamefont {Algora}, \citenamefont
  {Al-Khatib}, \citenamefont {Bastin}, \citenamefont {Benzoni}, \citenamefont
  {Borcea}, \citenamefont {Bourgeois}, \citenamefont {Bringel}, \citenamefont
  {Cl\'ement}, \citenamefont {Dalouzy}, \citenamefont {Dlouh\'y}, \citenamefont
  {Dombr\'adi}, \citenamefont {Drouart}, \citenamefont {Engelhardt},
  \citenamefont {Franchoo}, \citenamefont {F\"ul\"op}, \citenamefont
  {G\"orgen}, \citenamefont {Gr\'evy}, \citenamefont {H\"ubel}, \citenamefont
  {Ibrahim}, \citenamefont {Korten}, \citenamefont {Mr\'azek}, \citenamefont
  {Navin}, \citenamefont {Rotaru}, \citenamefont {Roussel~Chomaz},
  \citenamefont {Saint-Laurent}, \citenamefont {Sletten}, \citenamefont
  {Sohler}, \citenamefont {Sorlin}, \citenamefont {Stanoiu}, \citenamefont
  {Stefan}, \citenamefont {Theisen}, \citenamefont {Timis}, \citenamefont
  {Verney},\ and\ \citenamefont {Williams}}]{Bur12a}%
  \BibitemOpen
  \bibfield  {author} {\bibinfo {author} {\bibfnamefont {A.}~\bibnamefont
  {B\"urger}}, \bibinfo {author} {\bibfnamefont {F.}~\bibnamefont {Azaiez}},
  \bibinfo {author} {\bibfnamefont {A.}~\bibnamefont {Algora}}, \bibinfo
  {author} {\bibfnamefont {A.}~\bibnamefont {Al-Khatib}}, \bibinfo {author}
  {\bibfnamefont {B.}~\bibnamefont {Bastin}}, \bibinfo {author} {\bibfnamefont
  {G.}~\bibnamefont {Benzoni}}, \bibinfo {author} {\bibfnamefont
  {R.}~\bibnamefont {Borcea}}, \bibinfo {author} {\bibfnamefont
  {C.}~\bibnamefont {Bourgeois}}, \bibinfo {author} {\bibfnamefont
  {P.}~\bibnamefont {Bringel}}, \bibinfo {author} {\bibfnamefont
  {E.}~\bibnamefont {Cl\'ement}}, \bibinfo {author} {\bibfnamefont {J.-C.}\
  \bibnamefont {Dalouzy}}, \bibinfo {author} {\bibfnamefont {Z.}~\bibnamefont
  {Dlouh\'y}}, \bibinfo {author} {\bibfnamefont {Z.}~\bibnamefont
  {Dombr\'adi}}, \bibinfo {author} {\bibfnamefont {A.}~\bibnamefont {Drouart}},
  \bibinfo {author} {\bibfnamefont {C.}~\bibnamefont {Engelhardt}}, \bibinfo
  {author} {\bibfnamefont {S.}~\bibnamefont {Franchoo}}, \bibinfo {author}
  {\bibfnamefont {Z.}~\bibnamefont {F\"ul\"op}}, \bibinfo {author}
  {\bibfnamefont {A.}~\bibnamefont {G\"orgen}}, \bibinfo {author}
  {\bibfnamefont {S.}~\bibnamefont {Gr\'evy}}, \bibinfo {author} {\bibfnamefont
  {H.}~\bibnamefont {H\"ubel}}, \bibinfo {author} {\bibfnamefont
  {F.}~\bibnamefont {Ibrahim}}, \bibinfo {author} {\bibfnamefont
  {W.}~\bibnamefont {Korten}}, \bibinfo {author} {\bibfnamefont
  {J.}~\bibnamefont {Mr\'azek}}, \bibinfo {author} {\bibfnamefont
  {A.}~\bibnamefont {Navin}}, \bibinfo {author} {\bibfnamefont
  {F.}~\bibnamefont {Rotaru}}, \bibinfo {author} {\bibfnamefont
  {P.}~\bibnamefont {Roussel~Chomaz}}, \bibinfo {author} {\bibfnamefont
  {M.-G.}\ \bibnamefont {Saint-Laurent}}, \bibinfo {author} {\bibfnamefont
  {G.}~\bibnamefont {Sletten}}, \bibinfo {author} {\bibfnamefont
  {D.}~\bibnamefont {Sohler}}, \bibinfo {author} {\bibfnamefont
  {O.}~\bibnamefont {Sorlin}}, \bibinfo {author} {\bibfnamefont
  {M.}~\bibnamefont {Stanoiu}}, \bibinfo {author} {\bibfnamefont
  {I.}~\bibnamefont {Stefan}}, \bibinfo {author} {\bibfnamefont
  {C.}~\bibnamefont {Theisen}}, \bibinfo {author} {\bibfnamefont
  {C.}~\bibnamefont {Timis}}, \bibinfo {author} {\bibfnamefont
  {D.}~\bibnamefont {Verney}},\ and\ \bibinfo {author} {\bibfnamefont
  {S.}~\bibnamefont {Williams}},\ }\bibfield  {title} {\bibinfo {title} {Cross
  sections for one-neutron knock-out from ${}^{37}\mathrm{Ca}$ at intermediate
  energy},\ }\href {https://doi.org/10.1103/PhysRevC.86.064609} {\bibfield
  {journal} {\bibinfo  {journal} {Phys. Rev. C}\ }\textbf {\bibinfo {volume}
  {86}},\ \bibinfo {pages} {064609} (\bibinfo {year} {2012})}\BibitemShut
  {NoStop}%
\bibitem [{\citenamefont {Agostinelli}\ \emph {et~al.}(2003)\citenamefont
  {Agostinelli}, \citenamefont {Allison}, \citenamefont {Amako}, \citenamefont
  {Apostolakis}, \citenamefont {Araujo}, \citenamefont {Arce}, \citenamefont
  {Asai}, \citenamefont {Axen}, \citenamefont {Banerjee}, \citenamefont
  {Barrand}, \citenamefont {Behner}, \citenamefont {Bellagamba}, \citenamefont
  {Boudreau}, \citenamefont {Broglia}, \citenamefont {Brunengo}, \citenamefont
  {Burkhardt}, \citenamefont {Chauvie}, \citenamefont {Chuma}, \citenamefont
  {Chytracek}, \citenamefont {Cooperman}, \citenamefont {Cosmo}, \citenamefont
  {Degtyarenko}, \citenamefont {Dell'Acqua}, \citenamefont {Depaola},
  \citenamefont {Dietrich}, \citenamefont {Enami}, \citenamefont {Feliciello},
  \citenamefont {Ferguson}, \citenamefont {Fesefeldt}, \citenamefont {Folger},
  \citenamefont {Foppiano}, \citenamefont {Forti}, \citenamefont {Garelli},
  \citenamefont {Giani}, \citenamefont {Giannitrapani}, \citenamefont {Gibin},
  \citenamefont {Cadenas}, \citenamefont {Gonz{\'a}lez}, \citenamefont {Abril},
  \citenamefont {Greeniaus}, \citenamefont {Greiner}, \citenamefont {Grichine},
  \citenamefont {Grossheim}, \citenamefont {Guatelli}, \citenamefont
  {Gumplinger}, \citenamefont {Hamatsu}, \citenamefont {Hashimoto},
  \citenamefont {Hasui}, \citenamefont {Heikkinen}, \citenamefont {Howard},
  \citenamefont {Ivanchenko}, \citenamefont {Johnson}, \citenamefont {Jones},
  \citenamefont {Kallenbach}, \citenamefont {Kanaya}, \citenamefont {Kawabata},
  \citenamefont {Kawabata}, \citenamefont {Kawaguti}, \citenamefont {Kelner},
  \citenamefont {Kent}, \citenamefont {Kimura}, \citenamefont {Kodama},
  \citenamefont {Kokoulin}, \citenamefont {Kossov}, \citenamefont {Kurashige},
  \citenamefont {Lamanna}, \citenamefont {Lamp{\'e}n}, \citenamefont {Lara},
  \citenamefont {Lefebure}, \citenamefont {Lei}, \citenamefont {Liendl},
  \citenamefont {Lockman}, \citenamefont {Longo}, \citenamefont {Magni},
  \citenamefont {Maire}, \citenamefont {Medernach}, \citenamefont {Minamimoto},
  \citenamefont {de~Freitas}, \citenamefont {Morita}, \citenamefont {Murakami},
  \citenamefont {Nagamatu}, \citenamefont {Nartallo}, \citenamefont {Nieminen},
  \citenamefont {Nishimura}, \citenamefont {Ohtsubo}, \citenamefont {Okamura},
  \citenamefont {O'Neale}, \citenamefont {Oohata}, \citenamefont {Paech},
  \citenamefont {Perl}, \citenamefont {Pfeiffer}, \citenamefont {Pia},
  \citenamefont {Ranjard}, \citenamefont {Rybin}, \citenamefont {Sadilov},
  \citenamefont {Salvo}, \citenamefont {Santin}, \citenamefont {Sasaki},
  \citenamefont {Savvas}, \citenamefont {Sawada}, \citenamefont {Scherer},
  \citenamefont {Sei}, \citenamefont {Sirotenko}, \citenamefont {Smith},
  \citenamefont {Starkov}, \citenamefont {Stoecker}, \citenamefont {Sulkimo},
  \citenamefont {Takahata}, \citenamefont {Tanaka}, \citenamefont {Tcherniaev},
  \citenamefont {Tehrani}, \citenamefont {Tropeano}, \citenamefont {Truscott},
  \citenamefont {Uno}, \citenamefont {Urban}, \citenamefont {Urban},
  \citenamefont {Verderi}, \citenamefont {Walkden}, \citenamefont {Wander},
  \citenamefont {Weber}, \citenamefont {Wellisch}, \citenamefont {Wenaus},
  \citenamefont {Williams}, \citenamefont {Wright}, \citenamefont {Yamada},
  \citenamefont {Yoshida},\ and\ \citenamefont {Zschiesche}}]{Ago03a}%
  \BibitemOpen
  \bibfield  {author} {\bibinfo {author} {\bibfnamefont {S.}~\bibnamefont
  {Agostinelli}}, \bibinfo {author} {\bibfnamefont {J.}~\bibnamefont
  {Allison}}, \bibinfo {author} {\bibfnamefont {K.}~\bibnamefont {Amako}},
  \bibinfo {author} {\bibfnamefont {J.}~\bibnamefont {Apostolakis}}, \bibinfo
  {author} {\bibfnamefont {H.}~\bibnamefont {Araujo}}, \bibinfo {author}
  {\bibfnamefont {P.}~\bibnamefont {Arce}}, \bibinfo {author} {\bibfnamefont
  {M.}~\bibnamefont {Asai}}, \bibinfo {author} {\bibfnamefont {D.}~\bibnamefont
  {Axen}}, \bibinfo {author} {\bibfnamefont {S.}~\bibnamefont {Banerjee}},
  \bibinfo {author} {\bibfnamefont {G.}~\bibnamefont {Barrand}}, \bibinfo
  {author} {\bibfnamefont {F.}~\bibnamefont {Behner}}, \bibinfo {author}
  {\bibfnamefont {L.}~\bibnamefont {Bellagamba}}, \bibinfo {author}
  {\bibfnamefont {J.}~\bibnamefont {Boudreau}}, \bibinfo {author}
  {\bibfnamefont {L.}~\bibnamefont {Broglia}}, \bibinfo {author} {\bibfnamefont
  {A.}~\bibnamefont {Brunengo}}, \bibinfo {author} {\bibfnamefont
  {H.}~\bibnamefont {Burkhardt}}, \bibinfo {author} {\bibfnamefont
  {S.}~\bibnamefont {Chauvie}}, \bibinfo {author} {\bibfnamefont
  {J.}~\bibnamefont {Chuma}}, \bibinfo {author} {\bibfnamefont
  {R.}~\bibnamefont {Chytracek}}, \bibinfo {author} {\bibfnamefont
  {G.}~\bibnamefont {Cooperman}}, \bibinfo {author} {\bibfnamefont
  {G.}~\bibnamefont {Cosmo}}, \bibinfo {author} {\bibfnamefont
  {P.}~\bibnamefont {Degtyarenko}}, \bibinfo {author} {\bibfnamefont
  {A.}~\bibnamefont {Dell'Acqua}}, \bibinfo {author} {\bibfnamefont
  {G.}~\bibnamefont {Depaola}}, \bibinfo {author} {\bibfnamefont
  {D.}~\bibnamefont {Dietrich}}, \bibinfo {author} {\bibfnamefont
  {R.}~\bibnamefont {Enami}}, \bibinfo {author} {\bibfnamefont
  {A.}~\bibnamefont {Feliciello}}, \bibinfo {author} {\bibfnamefont
  {C.}~\bibnamefont {Ferguson}}, \bibinfo {author} {\bibfnamefont
  {H.}~\bibnamefont {Fesefeldt}}, \bibinfo {author} {\bibfnamefont
  {G.}~\bibnamefont {Folger}}, \bibinfo {author} {\bibfnamefont
  {F.}~\bibnamefont {Foppiano}}, \bibinfo {author} {\bibfnamefont
  {A.}~\bibnamefont {Forti}}, \bibinfo {author} {\bibfnamefont
  {S.}~\bibnamefont {Garelli}}, \bibinfo {author} {\bibfnamefont
  {S.}~\bibnamefont {Giani}}, \bibinfo {author} {\bibfnamefont
  {R.}~\bibnamefont {Giannitrapani}}, \bibinfo {author} {\bibfnamefont
  {D.}~\bibnamefont {Gibin}}, \bibinfo {author} {\bibfnamefont {J.~G.}\
  \bibnamefont {Cadenas}}, \bibinfo {author} {\bibfnamefont {I.}~\bibnamefont
  {Gonz{\'a}lez}}, \bibinfo {author} {\bibfnamefont {G.~G.}\ \bibnamefont
  {Abril}}, \bibinfo {author} {\bibfnamefont {G.}~\bibnamefont {Greeniaus}},
  \bibinfo {author} {\bibfnamefont {W.}~\bibnamefont {Greiner}}, \bibinfo
  {author} {\bibfnamefont {V.}~\bibnamefont {Grichine}}, \bibinfo {author}
  {\bibfnamefont {A.}~\bibnamefont {Grossheim}}, \bibinfo {author}
  {\bibfnamefont {S.}~\bibnamefont {Guatelli}}, \bibinfo {author}
  {\bibfnamefont {P.}~\bibnamefont {Gumplinger}}, \bibinfo {author}
  {\bibfnamefont {R.}~\bibnamefont {Hamatsu}}, \bibinfo {author} {\bibfnamefont
  {K.}~\bibnamefont {Hashimoto}}, \bibinfo {author} {\bibfnamefont
  {H.}~\bibnamefont {Hasui}}, \bibinfo {author} {\bibfnamefont
  {A.}~\bibnamefont {Heikkinen}}, \bibinfo {author} {\bibfnamefont
  {A.}~\bibnamefont {Howard}}, \bibinfo {author} {\bibfnamefont
  {V.}~\bibnamefont {Ivanchenko}}, \bibinfo {author} {\bibfnamefont
  {A.}~\bibnamefont {Johnson}}, \bibinfo {author} {\bibfnamefont
  {F.}~\bibnamefont {Jones}}, \bibinfo {author} {\bibfnamefont
  {J.}~\bibnamefont {Kallenbach}}, \bibinfo {author} {\bibfnamefont
  {N.}~\bibnamefont {Kanaya}}, \bibinfo {author} {\bibfnamefont
  {M.}~\bibnamefont {Kawabata}}, \bibinfo {author} {\bibfnamefont
  {Y.}~\bibnamefont {Kawabata}}, \bibinfo {author} {\bibfnamefont
  {M.}~\bibnamefont {Kawaguti}}, \bibinfo {author} {\bibfnamefont
  {S.}~\bibnamefont {Kelner}}, \bibinfo {author} {\bibfnamefont
  {P.}~\bibnamefont {Kent}}, \bibinfo {author} {\bibfnamefont {A.}~\bibnamefont
  {Kimura}}, \bibinfo {author} {\bibfnamefont {T.}~\bibnamefont {Kodama}},
  \bibinfo {author} {\bibfnamefont {R.}~\bibnamefont {Kokoulin}}, \bibinfo
  {author} {\bibfnamefont {M.}~\bibnamefont {Kossov}}, \bibinfo {author}
  {\bibfnamefont {H.}~\bibnamefont {Kurashige}}, \bibinfo {author}
  {\bibfnamefont {E.}~\bibnamefont {Lamanna}}, \bibinfo {author} {\bibfnamefont
  {T.}~\bibnamefont {Lamp{\'e}n}}, \bibinfo {author} {\bibfnamefont
  {V.}~\bibnamefont {Lara}}, \bibinfo {author} {\bibfnamefont {V.}~\bibnamefont
  {Lefebure}}, \bibinfo {author} {\bibfnamefont {F.}~\bibnamefont {Lei}},
  \bibinfo {author} {\bibfnamefont {M.}~\bibnamefont {Liendl}}, \bibinfo
  {author} {\bibfnamefont {W.}~\bibnamefont {Lockman}}, \bibinfo {author}
  {\bibfnamefont {F.}~\bibnamefont {Longo}}, \bibinfo {author} {\bibfnamefont
  {S.}~\bibnamefont {Magni}}, \bibinfo {author} {\bibfnamefont
  {M.}~\bibnamefont {Maire}}, \bibinfo {author} {\bibfnamefont
  {E.}~\bibnamefont {Medernach}}, \bibinfo {author} {\bibfnamefont
  {K.}~\bibnamefont {Minamimoto}}, \bibinfo {author} {\bibfnamefont {P.~M.}\
  \bibnamefont {de~Freitas}}, \bibinfo {author} {\bibfnamefont
  {Y.}~\bibnamefont {Morita}}, \bibinfo {author} {\bibfnamefont
  {K.}~\bibnamefont {Murakami}}, \bibinfo {author} {\bibfnamefont
  {M.}~\bibnamefont {Nagamatu}}, \bibinfo {author} {\bibfnamefont
  {R.}~\bibnamefont {Nartallo}}, \bibinfo {author} {\bibfnamefont
  {P.}~\bibnamefont {Nieminen}}, \bibinfo {author} {\bibfnamefont
  {T.}~\bibnamefont {Nishimura}}, \bibinfo {author} {\bibfnamefont
  {K.}~\bibnamefont {Ohtsubo}}, \bibinfo {author} {\bibfnamefont
  {M.}~\bibnamefont {Okamura}}, \bibinfo {author} {\bibfnamefont
  {S.}~\bibnamefont {O'Neale}}, \bibinfo {author} {\bibfnamefont
  {Y.}~\bibnamefont {Oohata}}, \bibinfo {author} {\bibfnamefont
  {K.}~\bibnamefont {Paech}}, \bibinfo {author} {\bibfnamefont
  {J.}~\bibnamefont {Perl}}, \bibinfo {author} {\bibfnamefont {A.}~\bibnamefont
  {Pfeiffer}}, \bibinfo {author} {\bibfnamefont {M.}~\bibnamefont {Pia}},
  \bibinfo {author} {\bibfnamefont {F.}~\bibnamefont {Ranjard}}, \bibinfo
  {author} {\bibfnamefont {A.}~\bibnamefont {Rybin}}, \bibinfo {author}
  {\bibfnamefont {S.}~\bibnamefont {Sadilov}}, \bibinfo {author} {\bibfnamefont
  {E.~D.}\ \bibnamefont {Salvo}}, \bibinfo {author} {\bibfnamefont
  {G.}~\bibnamefont {Santin}}, \bibinfo {author} {\bibfnamefont
  {T.}~\bibnamefont {Sasaki}}, \bibinfo {author} {\bibfnamefont
  {N.}~\bibnamefont {Savvas}}, \bibinfo {author} {\bibfnamefont
  {Y.}~\bibnamefont {Sawada}}, \bibinfo {author} {\bibfnamefont
  {S.}~\bibnamefont {Scherer}}, \bibinfo {author} {\bibfnamefont
  {S.}~\bibnamefont {Sei}}, \bibinfo {author} {\bibfnamefont {V.}~\bibnamefont
  {Sirotenko}}, \bibinfo {author} {\bibfnamefont {D.}~\bibnamefont {Smith}},
  \bibinfo {author} {\bibfnamefont {N.}~\bibnamefont {Starkov}}, \bibinfo
  {author} {\bibfnamefont {H.}~\bibnamefont {Stoecker}}, \bibinfo {author}
  {\bibfnamefont {J.}~\bibnamefont {Sulkimo}}, \bibinfo {author} {\bibfnamefont
  {M.}~\bibnamefont {Takahata}}, \bibinfo {author} {\bibfnamefont
  {S.}~\bibnamefont {Tanaka}}, \bibinfo {author} {\bibfnamefont
  {E.}~\bibnamefont {Tcherniaev}}, \bibinfo {author} {\bibfnamefont {E.~S.}\
  \bibnamefont {Tehrani}}, \bibinfo {author} {\bibfnamefont {M.}~\bibnamefont
  {Tropeano}}, \bibinfo {author} {\bibfnamefont {P.}~\bibnamefont {Truscott}},
  \bibinfo {author} {\bibfnamefont {H.}~\bibnamefont {Uno}}, \bibinfo {author}
  {\bibfnamefont {L.}~\bibnamefont {Urban}}, \bibinfo {author} {\bibfnamefont
  {P.}~\bibnamefont {Urban}}, \bibinfo {author} {\bibfnamefont
  {M.}~\bibnamefont {Verderi}}, \bibinfo {author} {\bibfnamefont
  {A.}~\bibnamefont {Walkden}}, \bibinfo {author} {\bibfnamefont
  {W.}~\bibnamefont {Wander}}, \bibinfo {author} {\bibfnamefont
  {H.}~\bibnamefont {Weber}}, \bibinfo {author} {\bibfnamefont
  {J.}~\bibnamefont {Wellisch}}, \bibinfo {author} {\bibfnamefont
  {T.}~\bibnamefont {Wenaus}}, \bibinfo {author} {\bibfnamefont
  {D.}~\bibnamefont {Williams}}, \bibinfo {author} {\bibfnamefont
  {D.}~\bibnamefont {Wright}}, \bibinfo {author} {\bibfnamefont
  {T.}~\bibnamefont {Yamada}}, \bibinfo {author} {\bibfnamefont
  {H.}~\bibnamefont {Yoshida}},\ and\ \bibinfo {author} {\bibfnamefont
  {D.}~\bibnamefont {Zschiesche}},\ }\bibfield  {title} {\bibinfo {title}
  {{GEANT4} -- a simulation toolkit},\ }\href
  {https://doi.org/https://doi.org/10.1016/S0168-9002(03)01368-8} {\bibfield
  {journal} {\bibinfo  {journal} {Nucl. Instrum. Methods Phys. Res. A}\
  }\textbf {\bibinfo {volume} {506}},\ \bibinfo {pages} {250 } (\bibinfo {year}
  {2003})}\BibitemShut {NoStop}%
\bibitem [{\citenamefont {Allison}\ \emph {et~al.}(2016)\citenamefont
  {Allison}, \citenamefont {Amako}, \citenamefont {Apostolakis}, \citenamefont
  {Arce}, \citenamefont {Asai}, \citenamefont {Aso}, \citenamefont {Bagli},
  \citenamefont {Bagulya}, \citenamefont {Banerjee}, \citenamefont {Barrand},
  \citenamefont {Beck}, \citenamefont {Bogdanov}, \citenamefont {Brandt},
  \citenamefont {Brown}, \citenamefont {Burkhardt}, \citenamefont {Canal},
  \citenamefont {Cano-Ott}, \citenamefont {Chauvie}, \citenamefont {Cho},
  \citenamefont {Cirrone}, \citenamefont {Cooperman}, \citenamefont
  {Cort{\'e}s-Giraldo}, \citenamefont {Cosmo}, \citenamefont {Cuttone},
  \citenamefont {Depaola}, \citenamefont {Desorgher}, \citenamefont {Dong},
  \citenamefont {Dotti}, \citenamefont {Elvira}, \citenamefont {Folger},
  \citenamefont {Francis}, \citenamefont {Galoyan}, \citenamefont {Garnier},
  \citenamefont {Gayer}, \citenamefont {Genser}, \citenamefont {Grichine},
  \citenamefont {Guatelli}, \citenamefont {Gu{\`e}ye}, \citenamefont
  {Gumplinger}, \citenamefont {Howard}, \citenamefont {H{\v r}ivn{\'a}{\v
  c}ov{\'a}}, \citenamefont {Hwang}, \citenamefont {Incerti}, \citenamefont
  {Ivanchenko}, \citenamefont {Ivanchenko}, \citenamefont {Jones},
  \citenamefont {Jun}, \citenamefont {Kaitaniemi}, \citenamefont
  {Karakatsanis}, \citenamefont {Karamitros}, \citenamefont {Kelsey},
  \citenamefont {Kimura}, \citenamefont {Koi}, \citenamefont {Kurashige},
  \citenamefont {Lechner}, \citenamefont {Lee}, \citenamefont {Longo},
  \citenamefont {Maire}, \citenamefont {Mancusi}, \citenamefont {Mantero},
  \citenamefont {Mendoza}, \citenamefont {Morgan}, \citenamefont {Murakami},
  \citenamefont {Nikitina}, \citenamefont {Pandola}, \citenamefont {Paprocki},
  \citenamefont {Perl}, \citenamefont {Petrovi{\'c}}, \citenamefont {Pia},
  \citenamefont {Pokorski}, \citenamefont {Quesada}, \citenamefont {Raine},
  \citenamefont {Reis}, \citenamefont {Ribon}, \citenamefont {Fira},
  \citenamefont {Romano}, \citenamefont {Russo}, \citenamefont {Santin},
  \citenamefont {Sasaki}, \citenamefont {Sawkey}, \citenamefont {Shin},
  \citenamefont {Strakovsky}, \citenamefont {Taborda}, \citenamefont {Tanaka},
  \citenamefont {Tom{\'e}}, \citenamefont {Toshito}, \citenamefont {Tran},
  \citenamefont {Truscott}, \citenamefont {Urban}, \citenamefont {Uzhinsky},
  \citenamefont {Verbeke}, \citenamefont {Verderi}, \citenamefont {Wendt},
  \citenamefont {Wenzel}, \citenamefont {Wright}, \citenamefont {Wright},
  \citenamefont {Yamashita}, \citenamefont {Yarba},\ and\ \citenamefont
  {Yoshida}}]{All16a}%
  \BibitemOpen
  \bibfield  {author} {\bibinfo {author} {\bibfnamefont {J.}~\bibnamefont
  {Allison}}, \bibinfo {author} {\bibfnamefont {K.}~\bibnamefont {Amako}},
  \bibinfo {author} {\bibfnamefont {J.}~\bibnamefont {Apostolakis}}, \bibinfo
  {author} {\bibfnamefont {P.}~\bibnamefont {Arce}}, \bibinfo {author}
  {\bibfnamefont {M.}~\bibnamefont {Asai}}, \bibinfo {author} {\bibfnamefont
  {T.}~\bibnamefont {Aso}}, \bibinfo {author} {\bibfnamefont {E.}~\bibnamefont
  {Bagli}}, \bibinfo {author} {\bibfnamefont {A.}~\bibnamefont {Bagulya}},
  \bibinfo {author} {\bibfnamefont {S.}~\bibnamefont {Banerjee}}, \bibinfo
  {author} {\bibfnamefont {G.}~\bibnamefont {Barrand}}, \bibinfo {author}
  {\bibfnamefont {B.}~\bibnamefont {Beck}}, \bibinfo {author} {\bibfnamefont
  {A.}~\bibnamefont {Bogdanov}}, \bibinfo {author} {\bibfnamefont
  {D.}~\bibnamefont {Brandt}}, \bibinfo {author} {\bibfnamefont
  {J.}~\bibnamefont {Brown}}, \bibinfo {author} {\bibfnamefont
  {H.}~\bibnamefont {Burkhardt}}, \bibinfo {author} {\bibfnamefont
  {P.}~\bibnamefont {Canal}}, \bibinfo {author} {\bibfnamefont
  {D.}~\bibnamefont {Cano-Ott}}, \bibinfo {author} {\bibfnamefont
  {S.}~\bibnamefont {Chauvie}}, \bibinfo {author} {\bibfnamefont
  {K.}~\bibnamefont {Cho}}, \bibinfo {author} {\bibfnamefont {G.}~\bibnamefont
  {Cirrone}}, \bibinfo {author} {\bibfnamefont {G.}~\bibnamefont {Cooperman}},
  \bibinfo {author} {\bibfnamefont {M.}~\bibnamefont {Cort{\'e}s-Giraldo}},
  \bibinfo {author} {\bibfnamefont {G.}~\bibnamefont {Cosmo}}, \bibinfo
  {author} {\bibfnamefont {G.}~\bibnamefont {Cuttone}}, \bibinfo {author}
  {\bibfnamefont {G.}~\bibnamefont {Depaola}}, \bibinfo {author} {\bibfnamefont
  {L.}~\bibnamefont {Desorgher}}, \bibinfo {author} {\bibfnamefont
  {X.}~\bibnamefont {Dong}}, \bibinfo {author} {\bibfnamefont {A.}~\bibnamefont
  {Dotti}}, \bibinfo {author} {\bibfnamefont {V.}~\bibnamefont {Elvira}},
  \bibinfo {author} {\bibfnamefont {G.}~\bibnamefont {Folger}}, \bibinfo
  {author} {\bibfnamefont {Z.}~\bibnamefont {Francis}}, \bibinfo {author}
  {\bibfnamefont {A.}~\bibnamefont {Galoyan}}, \bibinfo {author} {\bibfnamefont
  {L.}~\bibnamefont {Garnier}}, \bibinfo {author} {\bibfnamefont
  {M.}~\bibnamefont {Gayer}}, \bibinfo {author} {\bibfnamefont
  {K.}~\bibnamefont {Genser}}, \bibinfo {author} {\bibfnamefont
  {V.}~\bibnamefont {Grichine}}, \bibinfo {author} {\bibfnamefont
  {S.}~\bibnamefont {Guatelli}}, \bibinfo {author} {\bibfnamefont
  {P.}~\bibnamefont {Gu{\`e}ye}}, \bibinfo {author} {\bibfnamefont
  {P.}~\bibnamefont {Gumplinger}}, \bibinfo {author} {\bibfnamefont
  {A.}~\bibnamefont {Howard}}, \bibinfo {author} {\bibfnamefont
  {I.}~\bibnamefont {H{\v r}ivn{\'a}{\v c}ov{\'a}}}, \bibinfo {author}
  {\bibfnamefont {S.}~\bibnamefont {Hwang}}, \bibinfo {author} {\bibfnamefont
  {S.}~\bibnamefont {Incerti}}, \bibinfo {author} {\bibfnamefont
  {A.}~\bibnamefont {Ivanchenko}}, \bibinfo {author} {\bibfnamefont
  {V.}~\bibnamefont {Ivanchenko}}, \bibinfo {author} {\bibfnamefont
  {F.}~\bibnamefont {Jones}}, \bibinfo {author} {\bibfnamefont
  {S.}~\bibnamefont {Jun}}, \bibinfo {author} {\bibfnamefont {P.}~\bibnamefont
  {Kaitaniemi}}, \bibinfo {author} {\bibfnamefont {N.}~\bibnamefont
  {Karakatsanis}}, \bibinfo {author} {\bibfnamefont {M.}~\bibnamefont
  {Karamitros}}, \bibinfo {author} {\bibfnamefont {M.}~\bibnamefont {Kelsey}},
  \bibinfo {author} {\bibfnamefont {A.}~\bibnamefont {Kimura}}, \bibinfo
  {author} {\bibfnamefont {T.}~\bibnamefont {Koi}}, \bibinfo {author}
  {\bibfnamefont {H.}~\bibnamefont {Kurashige}}, \bibinfo {author}
  {\bibfnamefont {A.}~\bibnamefont {Lechner}}, \bibinfo {author} {\bibfnamefont
  {S.}~\bibnamefont {Lee}}, \bibinfo {author} {\bibfnamefont {F.}~\bibnamefont
  {Longo}}, \bibinfo {author} {\bibfnamefont {M.}~\bibnamefont {Maire}},
  \bibinfo {author} {\bibfnamefont {D.}~\bibnamefont {Mancusi}}, \bibinfo
  {author} {\bibfnamefont {A.}~\bibnamefont {Mantero}}, \bibinfo {author}
  {\bibfnamefont {E.}~\bibnamefont {Mendoza}}, \bibinfo {author} {\bibfnamefont
  {B.}~\bibnamefont {Morgan}}, \bibinfo {author} {\bibfnamefont
  {K.}~\bibnamefont {Murakami}}, \bibinfo {author} {\bibfnamefont
  {T.}~\bibnamefont {Nikitina}}, \bibinfo {author} {\bibfnamefont
  {L.}~\bibnamefont {Pandola}}, \bibinfo {author} {\bibfnamefont
  {P.}~\bibnamefont {Paprocki}}, \bibinfo {author} {\bibfnamefont
  {J.}~\bibnamefont {Perl}}, \bibinfo {author} {\bibfnamefont {I.}~\bibnamefont
  {Petrovi{\'c}}}, \bibinfo {author} {\bibfnamefont {M.}~\bibnamefont {Pia}},
  \bibinfo {author} {\bibfnamefont {W.}~\bibnamefont {Pokorski}}, \bibinfo
  {author} {\bibfnamefont {J.}~\bibnamefont {Quesada}}, \bibinfo {author}
  {\bibfnamefont {M.}~\bibnamefont {Raine}}, \bibinfo {author} {\bibfnamefont
  {M.}~\bibnamefont {Reis}}, \bibinfo {author} {\bibfnamefont {A.}~\bibnamefont
  {Ribon}}, \bibinfo {author} {\bibfnamefont {A.~R.}\ \bibnamefont {Fira}},
  \bibinfo {author} {\bibfnamefont {F.}~\bibnamefont {Romano}}, \bibinfo
  {author} {\bibfnamefont {G.}~\bibnamefont {Russo}}, \bibinfo {author}
  {\bibfnamefont {G.}~\bibnamefont {Santin}}, \bibinfo {author} {\bibfnamefont
  {T.}~\bibnamefont {Sasaki}}, \bibinfo {author} {\bibfnamefont
  {D.}~\bibnamefont {Sawkey}}, \bibinfo {author} {\bibfnamefont
  {J.}~\bibnamefont {Shin}}, \bibinfo {author} {\bibfnamefont {I.}~\bibnamefont
  {Strakovsky}}, \bibinfo {author} {\bibfnamefont {A.}~\bibnamefont {Taborda}},
  \bibinfo {author} {\bibfnamefont {S.}~\bibnamefont {Tanaka}}, \bibinfo
  {author} {\bibfnamefont {B.}~\bibnamefont {Tom{\'e}}}, \bibinfo {author}
  {\bibfnamefont {T.}~\bibnamefont {Toshito}}, \bibinfo {author} {\bibfnamefont
  {H.}~\bibnamefont {Tran}}, \bibinfo {author} {\bibfnamefont {P.}~\bibnamefont
  {Truscott}}, \bibinfo {author} {\bibfnamefont {L.}~\bibnamefont {Urban}},
  \bibinfo {author} {\bibfnamefont {V.}~\bibnamefont {Uzhinsky}}, \bibinfo
  {author} {\bibfnamefont {J.}~\bibnamefont {Verbeke}}, \bibinfo {author}
  {\bibfnamefont {M.}~\bibnamefont {Verderi}}, \bibinfo {author} {\bibfnamefont
  {B.}~\bibnamefont {Wendt}}, \bibinfo {author} {\bibfnamefont
  {H.}~\bibnamefont {Wenzel}}, \bibinfo {author} {\bibfnamefont
  {D.}~\bibnamefont {Wright}}, \bibinfo {author} {\bibfnamefont
  {D.}~\bibnamefont {Wright}}, \bibinfo {author} {\bibfnamefont
  {T.}~\bibnamefont {Yamashita}}, \bibinfo {author} {\bibfnamefont
  {J.}~\bibnamefont {Yarba}},\ and\ \bibinfo {author} {\bibfnamefont
  {H.}~\bibnamefont {Yoshida}},\ }\bibfield  {title} {\bibinfo {title} {Recent
  developments in {GEANT}4},\ }\href
  {https://doi.org/https://doi.org/10.1016/j.nima.2016.06.125} {\bibfield
  {journal} {\bibinfo  {journal} {Nucl. Instrum. Methods Phys. Res. A}\
  }\textbf {\bibinfo {volume} {835}},\ \bibinfo {pages} {186 } (\bibinfo {year}
  {2016})}\BibitemShut {NoStop}%
\bibitem [{\citenamefont {Riley}\ \emph {et~al.}(2021)\citenamefont {Riley},
  \citenamefont {Weisshaar}, \citenamefont {Crawford}, \citenamefont
  {Agiorgousis}, \citenamefont {Campbell}, \citenamefont {Cromaz},
  \citenamefont {Fallon}, \citenamefont {Gade}, \citenamefont {Gregory},
  \citenamefont {Haldeman}, \citenamefont {Jarvis}, \citenamefont
  {Lawson-John}, \citenamefont {Roberts}, \citenamefont {Sadler},\ and\
  \citenamefont {Stine}}]{Ril21a}%
  \BibitemOpen
  \bibfield  {author} {\bibinfo {author} {\bibfnamefont {L.}~\bibnamefont
  {Riley}}, \bibinfo {author} {\bibfnamefont {D.}~\bibnamefont {Weisshaar}},
  \bibinfo {author} {\bibfnamefont {H.}~\bibnamefont {Crawford}}, \bibinfo
  {author} {\bibfnamefont {M.}~\bibnamefont {Agiorgousis}}, \bibinfo {author}
  {\bibfnamefont {C.}~\bibnamefont {Campbell}}, \bibinfo {author}
  {\bibfnamefont {M.}~\bibnamefont {Cromaz}}, \bibinfo {author} {\bibfnamefont
  {P.}~\bibnamefont {Fallon}}, \bibinfo {author} {\bibfnamefont
  {A.}~\bibnamefont {Gade}}, \bibinfo {author} {\bibfnamefont {S.}~\bibnamefont
  {Gregory}}, \bibinfo {author} {\bibfnamefont {E.}~\bibnamefont {Haldeman}},
  \bibinfo {author} {\bibfnamefont {L.}~\bibnamefont {Jarvis}}, \bibinfo
  {author} {\bibfnamefont {E.}~\bibnamefont {Lawson-John}}, \bibinfo {author}
  {\bibfnamefont {B.}~\bibnamefont {Roberts}}, \bibinfo {author} {\bibfnamefont
  {B.}~\bibnamefont {Sadler}},\ and\ \bibinfo {author} {\bibfnamefont
  {C.}~\bibnamefont {Stine}},\ }\bibfield  {title} {\bibinfo {title}
  {{UCGretina} {GEANT}4 simulation of the {GRETINA} {G}amma-{R}ay {E}nergy
  {T}racking {A}rray},\ }\href
  {https://doi.org/https://doi.org/10.1016/j.nima.2021.165305} {\bibfield
  {journal} {\bibinfo  {journal} {Nucl. Instrum. Methods Phys. Res. A}\
  }\textbf {\bibinfo {volume} {1003}},\ \bibinfo {pages} {165305} (\bibinfo
  {year} {2021})}\BibitemShut {NoStop}%
\bibitem [{\citenamefont {Bazin}\ \emph
  {et~al.}(2003{\natexlab{b}})\citenamefont {Bazin}, \citenamefont {Brown},
  \citenamefont {Campbell}, \citenamefont {Church}, \citenamefont {Dinca},
  \citenamefont {Enders}, \citenamefont {Gade}, \citenamefont {Glasmacher},
  \citenamefont {Hansen}, \citenamefont {Mueller}, \citenamefont {Olliver},
  \citenamefont {Perry}, \citenamefont {Sherrill}, \citenamefont {Terry},\ and\
  \citenamefont {Tostevin}}]{Baz03b}%
  \BibitemOpen
  \bibfield  {author} {\bibinfo {author} {\bibfnamefont {D.}~\bibnamefont
  {Bazin}}, \bibinfo {author} {\bibfnamefont {B.~A.}\ \bibnamefont {Brown}},
  \bibinfo {author} {\bibfnamefont {C.~M.}\ \bibnamefont {Campbell}}, \bibinfo
  {author} {\bibfnamefont {J.~A.}\ \bibnamefont {Church}}, \bibinfo {author}
  {\bibfnamefont {D.~C.}\ \bibnamefont {Dinca}}, \bibinfo {author}
  {\bibfnamefont {J.}~\bibnamefont {Enders}}, \bibinfo {author} {\bibfnamefont
  {A.}~\bibnamefont {Gade}}, \bibinfo {author} {\bibfnamefont {T.}~\bibnamefont
  {Glasmacher}}, \bibinfo {author} {\bibfnamefont {P.~G.}\ \bibnamefont
  {Hansen}}, \bibinfo {author} {\bibfnamefont {W.~F.}\ \bibnamefont {Mueller}},
  \bibinfo {author} {\bibfnamefont {H.}~\bibnamefont {Olliver}}, \bibinfo
  {author} {\bibfnamefont {B.~C.}\ \bibnamefont {Perry}}, \bibinfo {author}
  {\bibfnamefont {B.~M.}\ \bibnamefont {Sherrill}}, \bibinfo {author}
  {\bibfnamefont {J.~R.}\ \bibnamefont {Terry}},\ and\ \bibinfo {author}
  {\bibfnamefont {J.~A.}\ \bibnamefont {Tostevin}},\ }\bibfield  {title}
  {\bibinfo {title} {{N}ew {D}irect {R}eaction: {T}wo-{P}roton {K}nockout from
  {N}eutron-{R}ich {N}uclei},\ }\href
  {https://doi.org/10.1103/PhysRevLett.91.012501} {\bibfield  {journal}
  {\bibinfo  {journal} {Phys. Rev. Lett.}\ }\textbf {\bibinfo {volume} {91}},\
  \bibinfo {pages} {012501} (\bibinfo {year} {2003}{\natexlab{b}})}\BibitemShut
  {NoStop}%
\bibitem [{\citenamefont {Yoneda}\ \emph {et~al.}(2006)\citenamefont {Yoneda},
  \citenamefont {Obertelli}, \citenamefont {Gade}, \citenamefont {Bazin},
  \citenamefont {Brown}, \citenamefont {Campbell}, \citenamefont {Cook},
  \citenamefont {Cottle}, \citenamefont {Davies}, \citenamefont {Dinca},
  \citenamefont {Glasmacher}, \citenamefont {Hansen}, \citenamefont {Hoagland},
  \citenamefont {Kemper}, \citenamefont {Lecouey}, \citenamefont {Mueller},
  \citenamefont {Reynolds}, \citenamefont {Roeder}, \citenamefont {Terry},
  \citenamefont {Tostevin},\ and\ \citenamefont {Zwahlen}}]{Yon06a}%
  \BibitemOpen
  \bibfield  {author} {\bibinfo {author} {\bibfnamefont {K.}~\bibnamefont
  {Yoneda}}, \bibinfo {author} {\bibfnamefont {A.}~\bibnamefont {Obertelli}},
  \bibinfo {author} {\bibfnamefont {A.}~\bibnamefont {Gade}}, \bibinfo {author}
  {\bibfnamefont {D.}~\bibnamefont {Bazin}}, \bibinfo {author} {\bibfnamefont
  {B.~A.}\ \bibnamefont {Brown}}, \bibinfo {author} {\bibfnamefont {C.~M.}\
  \bibnamefont {Campbell}}, \bibinfo {author} {\bibfnamefont {J.~M.}\
  \bibnamefont {Cook}}, \bibinfo {author} {\bibfnamefont {P.~D.}\ \bibnamefont
  {Cottle}}, \bibinfo {author} {\bibfnamefont {A.~D.}\ \bibnamefont {Davies}},
  \bibinfo {author} {\bibfnamefont {D.-C.}\ \bibnamefont {Dinca}}, \bibinfo
  {author} {\bibfnamefont {T.}~\bibnamefont {Glasmacher}}, \bibinfo {author}
  {\bibfnamefont {P.~G.}\ \bibnamefont {Hansen}}, \bibinfo {author}
  {\bibfnamefont {T.}~\bibnamefont {Hoagland}}, \bibinfo {author}
  {\bibfnamefont {K.~W.}\ \bibnamefont {Kemper}}, \bibinfo {author}
  {\bibfnamefont {J.-L.}\ \bibnamefont {Lecouey}}, \bibinfo {author}
  {\bibfnamefont {W.~F.}\ \bibnamefont {Mueller}}, \bibinfo {author}
  {\bibfnamefont {R.~R.}\ \bibnamefont {Reynolds}}, \bibinfo {author}
  {\bibfnamefont {B.~T.}\ \bibnamefont {Roeder}}, \bibinfo {author}
  {\bibfnamefont {J.~R.}\ \bibnamefont {Terry}}, \bibinfo {author}
  {\bibfnamefont {J.~A.}\ \bibnamefont {Tostevin}},\ and\ \bibinfo {author}
  {\bibfnamefont {H.}~\bibnamefont {Zwahlen}},\ }\bibfield  {title} {\bibinfo
  {title} {Two-neutron knockout from neutron-deficient $^{34}\mathrm{Ar}$,
  $^{30}\mathrm{S}$, and $^{26}\mathrm{Si}$},\ }\href
  {https://doi.org/10.1103/PhysRevC.74.021303} {\bibfield  {journal} {\bibinfo
  {journal} {Phys. Rev. C}\ }\textbf {\bibinfo {volume} {74}},\ \bibinfo
  {pages} {021303} (\bibinfo {year} {2006})}\BibitemShut {NoStop}%
\bibitem [{\citenamefont {Simpson}\ \emph
  {et~al.}(2009{\natexlab{a}})\citenamefont {Simpson}, \citenamefont
  {Tostevin}, \citenamefont {Bazin}, \citenamefont {Brown},\ and\ \citenamefont
  {Gade}}]{Sim09a}%
  \BibitemOpen
  \bibfield  {author} {\bibinfo {author} {\bibfnamefont {E.~C.}\ \bibnamefont
  {Simpson}}, \bibinfo {author} {\bibfnamefont {J.~A.}\ \bibnamefont
  {Tostevin}}, \bibinfo {author} {\bibfnamefont {D.}~\bibnamefont {Bazin}},
  \bibinfo {author} {\bibfnamefont {B.~A.}\ \bibnamefont {Brown}},\ and\
  \bibinfo {author} {\bibfnamefont {A.}~\bibnamefont {Gade}},\ }\bibfield
  {title} {\bibinfo {title} {{T}wo-{N}ucleon {K}nockout {S}pectroscopy at the
  {L}imits of {N}uclear {S}tability},\ }\href
  {https://doi.org/10.1103/PhysRevLett.102.132502} {\bibfield  {journal}
  {\bibinfo  {journal} {Phys. Rev. Lett.}\ }\textbf {\bibinfo {volume} {102}},\
  \bibinfo {pages} {132502} (\bibinfo {year} {2009}{\natexlab{a}})}\BibitemShut
  {NoStop}%
\bibitem [{\citenamefont {Simpson}\ \emph
  {et~al.}(2009{\natexlab{b}})\citenamefont {Simpson}, \citenamefont
  {Tostevin}, \citenamefont {Bazin},\ and\ \citenamefont {Gade}}]{Sim09b}%
  \BibitemOpen
  \bibfield  {author} {\bibinfo {author} {\bibfnamefont {E.~C.}\ \bibnamefont
  {Simpson}}, \bibinfo {author} {\bibfnamefont {J.~A.}\ \bibnamefont
  {Tostevin}}, \bibinfo {author} {\bibfnamefont {D.}~\bibnamefont {Bazin}},\
  and\ \bibinfo {author} {\bibfnamefont {A.}~\bibnamefont {Gade}},\ }\bibfield
  {title} {\bibinfo {title} {Longitudinal momentum distributions of the
  reaction residues following fast two-nucleon knockout reactions},\ }\href
  {https://doi.org/10.1103/PhysRevC.79.064621} {\bibfield  {journal} {\bibinfo
  {journal} {Phys. Rev. C}\ }\textbf {\bibinfo {volume} {79}},\ \bibinfo
  {pages} {064621} (\bibinfo {year} {2009}{\natexlab{b}})}\BibitemShut
  {NoStop}%
\bibitem [{\citenamefont {Simpson}\ and\ \citenamefont
  {Tostevin}(2010)}]{Sim10a}%
  \BibitemOpen
  \bibfield  {author} {\bibinfo {author} {\bibfnamefont {E.~C.}\ \bibnamefont
  {Simpson}}\ and\ \bibinfo {author} {\bibfnamefont {J.~A.}\ \bibnamefont
  {Tostevin}},\ }\bibfield  {title} {\bibinfo {title} {Correlations probed in
  direct two-nucleon removal reactions},\ }\href
  {https://doi.org/10.1103/PhysRevC.82.044616} {\bibfield  {journal} {\bibinfo
  {journal} {Phys. Rev. C}\ }\textbf {\bibinfo {volume} {82}},\ \bibinfo
  {pages} {044616} (\bibinfo {year} {2010})}\BibitemShut {NoStop}%
\bibitem [{\citenamefont {Longfellow}\ \emph {et~al.}(2020)\citenamefont
  {Longfellow}, \citenamefont {Gade}, \citenamefont {Tostevin}, \citenamefont
  {Simpson}, \citenamefont {Brown}, \citenamefont {Magilligan}, \citenamefont
  {Bazin}, \citenamefont {Bender}, \citenamefont {Bowry}, \citenamefont
  {Elman}, \citenamefont {Lunderberg}, \citenamefont {Rhodes}, \citenamefont
  {Spieker}, \citenamefont {Weisshaar},\ and\ \citenamefont
  {Williams}}]{Lon20a}%
  \BibitemOpen
  \bibfield  {author} {\bibinfo {author} {\bibfnamefont {B.}~\bibnamefont
  {Longfellow}}, \bibinfo {author} {\bibfnamefont {A.}~\bibnamefont {Gade}},
  \bibinfo {author} {\bibfnamefont {J.~A.}\ \bibnamefont {Tostevin}}, \bibinfo
  {author} {\bibfnamefont {E.~C.}\ \bibnamefont {Simpson}}, \bibinfo {author}
  {\bibfnamefont {B.~A.}\ \bibnamefont {Brown}}, \bibinfo {author}
  {\bibfnamefont {A.}~\bibnamefont {Magilligan}}, \bibinfo {author}
  {\bibfnamefont {D.}~\bibnamefont {Bazin}}, \bibinfo {author} {\bibfnamefont
  {P.~C.}\ \bibnamefont {Bender}}, \bibinfo {author} {\bibfnamefont
  {M.}~\bibnamefont {Bowry}}, \bibinfo {author} {\bibfnamefont
  {B.}~\bibnamefont {Elman}}, \bibinfo {author} {\bibfnamefont
  {E.}~\bibnamefont {Lunderberg}}, \bibinfo {author} {\bibfnamefont
  {D.}~\bibnamefont {Rhodes}}, \bibinfo {author} {\bibfnamefont
  {M.}~\bibnamefont {Spieker}}, \bibinfo {author} {\bibfnamefont
  {D.}~\bibnamefont {Weisshaar}},\ and\ \bibinfo {author} {\bibfnamefont
  {S.~J.}\ \bibnamefont {Williams}},\ }\bibfield  {title} {\bibinfo {title}
  {Two-neutron knockout as a probe of the composition of states in
  $^{22}\mathrm{Mg},^{23}\mathrm{Al}$, and $^{24}\mathrm{Si}$},\ }\href
  {https://doi.org/10.1103/PhysRevC.101.031303} {\bibfield  {journal} {\bibinfo
   {journal} {Phys. Rev. C}\ }\textbf {\bibinfo {volume} {101}},\ \bibinfo
  {pages} {031303} (\bibinfo {year} {2020})}\BibitemShut {NoStop}%
\bibitem [{\citenamefont {Stroberg}\ \emph {et~al.}(2014)\citenamefont
  {Stroberg}, \citenamefont {Gade}, \citenamefont {Tostevin}, \citenamefont
  {Bader}, \citenamefont {Baugher}, \citenamefont {Bazin}, \citenamefont
  {Berryman}, \citenamefont {Brown}, \citenamefont {Campbell}, \citenamefont
  {Kemper}, \citenamefont {Langer}, \citenamefont {Lunderberg}, \citenamefont
  {Lemasson}, \citenamefont {Noji}, \citenamefont {Recchia}, \citenamefont
  {Walz}, \citenamefont {Weisshaar},\ and\ \citenamefont {Williams}}]{Str14a}%
  \BibitemOpen
  \bibfield  {author} {\bibinfo {author} {\bibfnamefont {S.~R.}\ \bibnamefont
  {Stroberg}}, \bibinfo {author} {\bibfnamefont {A.}~\bibnamefont {Gade}},
  \bibinfo {author} {\bibfnamefont {J.~A.}\ \bibnamefont {Tostevin}}, \bibinfo
  {author} {\bibfnamefont {V.~M.}\ \bibnamefont {Bader}}, \bibinfo {author}
  {\bibfnamefont {T.}~\bibnamefont {Baugher}}, \bibinfo {author} {\bibfnamefont
  {D.}~\bibnamefont {Bazin}}, \bibinfo {author} {\bibfnamefont {J.~S.}\
  \bibnamefont {Berryman}}, \bibinfo {author} {\bibfnamefont {B.~A.}\
  \bibnamefont {Brown}}, \bibinfo {author} {\bibfnamefont {C.~M.}\ \bibnamefont
  {Campbell}}, \bibinfo {author} {\bibfnamefont {K.~W.}\ \bibnamefont
  {Kemper}}, \bibinfo {author} {\bibfnamefont {C.}~\bibnamefont {Langer}},
  \bibinfo {author} {\bibfnamefont {E.}~\bibnamefont {Lunderberg}}, \bibinfo
  {author} {\bibfnamefont {A.}~\bibnamefont {Lemasson}}, \bibinfo {author}
  {\bibfnamefont {S.}~\bibnamefont {Noji}}, \bibinfo {author} {\bibfnamefont
  {F.}~\bibnamefont {Recchia}}, \bibinfo {author} {\bibfnamefont
  {C.}~\bibnamefont {Walz}}, \bibinfo {author} {\bibfnamefont {D.}~\bibnamefont
  {Weisshaar}},\ and\ \bibinfo {author} {\bibfnamefont {S.~J.}\ \bibnamefont
  {Williams}},\ }\bibfield  {title} {\bibinfo {title} {Single-particle
  structure of silicon isotopes approaching $^{42}\mathrm{Si}$},\ }\href
  {https://doi.org/10.1103/PhysRevC.90.034301} {\bibfield  {journal} {\bibinfo
  {journal} {Phys. Rev. C}\ }\textbf {\bibinfo {volume} {90}},\ \bibinfo
  {pages} {034301} (\bibinfo {year} {2014})}\BibitemShut {NoStop}%
\bibitem [{\citenamefont {Magilligan}\ and\ \citenamefont
  {Brown}(2020)}]{Mag20a}%
  \BibitemOpen
  \bibfield  {author} {\bibinfo {author} {\bibfnamefont {A.}~\bibnamefont
  {Magilligan}}\ and\ \bibinfo {author} {\bibfnamefont {B.~A.}\ \bibnamefont
  {Brown}},\ }\bibfield  {title} {\bibinfo {title} {New isospin-breaking
  ``{USD}'' {H}amiltonians for the $\mathit{sd}$ shell},\ }\href
  {https://doi.org/10.1103/PhysRevC.101.064312} {\bibfield  {journal} {\bibinfo
   {journal} {Phys. Rev. C}\ }\textbf {\bibinfo {volume} {101}},\ \bibinfo
  {pages} {064312} (\bibinfo {year} {2020})}\BibitemShut {NoStop}%
\bibitem [{\citenamefont {Gade}\ \emph {et~al.}(2007)\citenamefont {Gade},
  \citenamefont {Adrich}, \citenamefont {Bazin}, \citenamefont {Bowen},
  \citenamefont {Brown}, \citenamefont {Campbell}, \citenamefont {Cook},
  \citenamefont {Ettenauer}, \citenamefont {Glasmacher}, \citenamefont
  {Kemper}, \citenamefont {McDaniel}, \citenamefont {Obertelli}, \citenamefont
  {Otsuka}, \citenamefont {Ratkiewicz}, \citenamefont {Siwek}, \citenamefont
  {Terry}, \citenamefont {Tostevin}, \citenamefont {Utsuno},\ and\
  \citenamefont {Weisshaar}}]{Gad07a}%
  \BibitemOpen
  \bibfield  {author} {\bibinfo {author} {\bibfnamefont {A.}~\bibnamefont
  {Gade}}, \bibinfo {author} {\bibfnamefont {P.}~\bibnamefont {Adrich}},
  \bibinfo {author} {\bibfnamefont {D.}~\bibnamefont {Bazin}}, \bibinfo
  {author} {\bibfnamefont {M.~D.}\ \bibnamefont {Bowen}}, \bibinfo {author}
  {\bibfnamefont {B.~A.}\ \bibnamefont {Brown}}, \bibinfo {author}
  {\bibfnamefont {C.~M.}\ \bibnamefont {Campbell}}, \bibinfo {author}
  {\bibfnamefont {J.~M.}\ \bibnamefont {Cook}}, \bibinfo {author}
  {\bibfnamefont {S.}~\bibnamefont {Ettenauer}}, \bibinfo {author}
  {\bibfnamefont {T.}~\bibnamefont {Glasmacher}}, \bibinfo {author}
  {\bibfnamefont {K.~W.}\ \bibnamefont {Kemper}}, \bibinfo {author}
  {\bibfnamefont {S.}~\bibnamefont {McDaniel}}, \bibinfo {author}
  {\bibfnamefont {A.}~\bibnamefont {Obertelli}}, \bibinfo {author}
  {\bibfnamefont {T.}~\bibnamefont {Otsuka}}, \bibinfo {author} {\bibfnamefont
  {A.}~\bibnamefont {Ratkiewicz}}, \bibinfo {author} {\bibfnamefont
  {K.}~\bibnamefont {Siwek}}, \bibinfo {author} {\bibfnamefont {J.~R.}\
  \bibnamefont {Terry}}, \bibinfo {author} {\bibfnamefont {J.~A.}\ \bibnamefont
  {Tostevin}}, \bibinfo {author} {\bibfnamefont {Y.}~\bibnamefont {Utsuno}},\
  and\ \bibinfo {author} {\bibfnamefont {D.}~\bibnamefont {Weisshaar}},\
  }\bibfield  {title} {\bibinfo {title} {{S}pectroscopy of $^{36}\mathrm{Mg}$:
  {I}nterplay of {N}ormal and {I}ntruder {C}onfigurations at the
  {N}eutron-{R}ich {B}oundary of the ``{I}sland of {I}nversion''},\ }\href
  {https://doi.org/10.1103/PhysRevLett.99.072502} {\bibfield  {journal}
  {\bibinfo  {journal} {Phys. Rev. Lett.}\ }\textbf {\bibinfo {volume} {99}},\
  \bibinfo {pages} {072502} (\bibinfo {year} {2007})}\BibitemShut {NoStop}%
\bibitem [{\citenamefont {Fallon}\ \emph {et~al.}(2010)\citenamefont {Fallon},
  \citenamefont {Rodriguez-Vieitez}, \citenamefont {Macchiavelli},
  \citenamefont {Gade}, \citenamefont {Tostevin}, \citenamefont {Adrich},
  \citenamefont {Bazin}, \citenamefont {Bowen}, \citenamefont {Campbell},
  \citenamefont {Clark}, \citenamefont {Cook}, \citenamefont {Cromaz},
  \citenamefont {Dinca}, \citenamefont {Glasmacher}, \citenamefont {Lee},
  \citenamefont {McDaniel}, \citenamefont {Mueller}, \citenamefont {Prussin},
  \citenamefont {Ratkiewicz}, \citenamefont {Siwek}, \citenamefont {Terry},
  \citenamefont {Weisshaar}, \citenamefont {Wiedeking}, \citenamefont {Yoneda},
  \citenamefont {Brown}, \citenamefont {Otsuka},\ and\ \citenamefont
  {Utsuno}}]{Fal10a}%
  \BibitemOpen
  \bibfield  {author} {\bibinfo {author} {\bibfnamefont {P.}~\bibnamefont
  {Fallon}}, \bibinfo {author} {\bibfnamefont {E.}~\bibnamefont
  {Rodriguez-Vieitez}}, \bibinfo {author} {\bibfnamefont {A.~O.}\ \bibnamefont
  {Macchiavelli}}, \bibinfo {author} {\bibfnamefont {A.}~\bibnamefont {Gade}},
  \bibinfo {author} {\bibfnamefont {J.~A.}\ \bibnamefont {Tostevin}}, \bibinfo
  {author} {\bibfnamefont {P.}~\bibnamefont {Adrich}}, \bibinfo {author}
  {\bibfnamefont {D.}~\bibnamefont {Bazin}}, \bibinfo {author} {\bibfnamefont
  {M.}~\bibnamefont {Bowen}}, \bibinfo {author} {\bibfnamefont {C.~M.}\
  \bibnamefont {Campbell}}, \bibinfo {author} {\bibfnamefont {R.~M.}\
  \bibnamefont {Clark}}, \bibinfo {author} {\bibfnamefont {J.~M.}\ \bibnamefont
  {Cook}}, \bibinfo {author} {\bibfnamefont {M.}~\bibnamefont {Cromaz}},
  \bibinfo {author} {\bibfnamefont {D.~C.}\ \bibnamefont {Dinca}}, \bibinfo
  {author} {\bibfnamefont {T.}~\bibnamefont {Glasmacher}}, \bibinfo {author}
  {\bibfnamefont {I.~Y.}\ \bibnamefont {Lee}}, \bibinfo {author} {\bibfnamefont
  {S.}~\bibnamefont {McDaniel}}, \bibinfo {author} {\bibfnamefont {W.~F.}\
  \bibnamefont {Mueller}}, \bibinfo {author} {\bibfnamefont {S.~G.}\
  \bibnamefont {Prussin}}, \bibinfo {author} {\bibfnamefont {A.}~\bibnamefont
  {Ratkiewicz}}, \bibinfo {author} {\bibfnamefont {K.}~\bibnamefont {Siwek}},
  \bibinfo {author} {\bibfnamefont {J.~R.}\ \bibnamefont {Terry}}, \bibinfo
  {author} {\bibfnamefont {D.}~\bibnamefont {Weisshaar}}, \bibinfo {author}
  {\bibfnamefont {M.}~\bibnamefont {Wiedeking}}, \bibinfo {author}
  {\bibfnamefont {K.}~\bibnamefont {Yoneda}}, \bibinfo {author} {\bibfnamefont
  {B.~A.}\ \bibnamefont {Brown}}, \bibinfo {author} {\bibfnamefont
  {T.}~\bibnamefont {Otsuka}},\ and\ \bibinfo {author} {\bibfnamefont
  {Y.}~\bibnamefont {Utsuno}},\ }\bibfield  {title} {\bibinfo {title}
  {{T}wo-proton knockout from $^{32}\mathrm{Mg}$: {I}ntruder amplitudes in
  $^{30}\mathrm{Ne}$ and implications for the binding of
  $^{29,31}\mathrm{F}$},\ }\href {https://doi.org/10.1103/PhysRevC.81.041302}
  {\bibfield  {journal} {\bibinfo  {journal} {Phys. Rev. C}\ }\textbf {\bibinfo
  {volume} {81}},\ \bibinfo {pages} {041302} (\bibinfo {year}
  {2010})}\BibitemShut {NoStop}%
\bibitem [{\citenamefont {Crawford}\ \emph {et~al.}(2014)\citenamefont
  {Crawford}, \citenamefont {Fallon}, \citenamefont {Macchiavelli},
  \citenamefont {Clark}, \citenamefont {Brown}, \citenamefont {Tostevin},
  \citenamefont {Bazin}, \citenamefont {Aoi}, \citenamefont {Doornenbal},
  \citenamefont {Matsushita}, \citenamefont {Scheit}, \citenamefont
  {Steppenbeck}, \citenamefont {Takeuchi}, \citenamefont {Baba}, \citenamefont
  {Campbell}, \citenamefont {Cromaz}, \citenamefont {Ideguchi}, \citenamefont
  {Kobayashi}, \citenamefont {Kondo}, \citenamefont {Lee}, \citenamefont {Lee},
  \citenamefont {Lee}, \citenamefont {Li}, \citenamefont {Michimasa},
  \citenamefont {Motobayashi}, \citenamefont {Nakamura}, \citenamefont {Ota},
  \citenamefont {Paschalis}, \citenamefont {Petri}, \citenamefont {Sako},
  \citenamefont {Sakurai}, \citenamefont {Shimoura}, \citenamefont {Takechi},
  \citenamefont {Togano}, \citenamefont {Wang},\ and\ \citenamefont
  {Yoneda}}]{Cra14a}%
  \BibitemOpen
  \bibfield  {author} {\bibinfo {author} {\bibfnamefont {H.~L.}\ \bibnamefont
  {Crawford}}, \bibinfo {author} {\bibfnamefont {P.}~\bibnamefont {Fallon}},
  \bibinfo {author} {\bibfnamefont {A.~O.}\ \bibnamefont {Macchiavelli}},
  \bibinfo {author} {\bibfnamefont {R.~M.}\ \bibnamefont {Clark}}, \bibinfo
  {author} {\bibfnamefont {B.~A.}\ \bibnamefont {Brown}}, \bibinfo {author}
  {\bibfnamefont {J.~A.}\ \bibnamefont {Tostevin}}, \bibinfo {author}
  {\bibfnamefont {D.}~\bibnamefont {Bazin}}, \bibinfo {author} {\bibfnamefont
  {N.}~\bibnamefont {Aoi}}, \bibinfo {author} {\bibfnamefont {P.}~\bibnamefont
  {Doornenbal}}, \bibinfo {author} {\bibfnamefont {M.}~\bibnamefont
  {Matsushita}}, \bibinfo {author} {\bibfnamefont {H.}~\bibnamefont {Scheit}},
  \bibinfo {author} {\bibfnamefont {D.}~\bibnamefont {Steppenbeck}}, \bibinfo
  {author} {\bibfnamefont {S.}~\bibnamefont {Takeuchi}}, \bibinfo {author}
  {\bibfnamefont {H.}~\bibnamefont {Baba}}, \bibinfo {author} {\bibfnamefont
  {C.~M.}\ \bibnamefont {Campbell}}, \bibinfo {author} {\bibfnamefont
  {M.}~\bibnamefont {Cromaz}}, \bibinfo {author} {\bibfnamefont
  {E.}~\bibnamefont {Ideguchi}}, \bibinfo {author} {\bibfnamefont
  {N.}~\bibnamefont {Kobayashi}}, \bibinfo {author} {\bibfnamefont
  {Y.}~\bibnamefont {Kondo}}, \bibinfo {author} {\bibfnamefont
  {G.}~\bibnamefont {Lee}}, \bibinfo {author} {\bibfnamefont {I.~Y.}\
  \bibnamefont {Lee}}, \bibinfo {author} {\bibfnamefont {J.}~\bibnamefont
  {Lee}}, \bibinfo {author} {\bibfnamefont {K.}~\bibnamefont {Li}}, \bibinfo
  {author} {\bibfnamefont {S.}~\bibnamefont {Michimasa}}, \bibinfo {author}
  {\bibfnamefont {T.}~\bibnamefont {Motobayashi}}, \bibinfo {author}
  {\bibfnamefont {T.}~\bibnamefont {Nakamura}}, \bibinfo {author}
  {\bibfnamefont {S.}~\bibnamefont {Ota}}, \bibinfo {author} {\bibfnamefont
  {S.}~\bibnamefont {Paschalis}}, \bibinfo {author} {\bibfnamefont
  {M.}~\bibnamefont {Petri}}, \bibinfo {author} {\bibfnamefont
  {T.}~\bibnamefont {Sako}}, \bibinfo {author} {\bibfnamefont {H.}~\bibnamefont
  {Sakurai}}, \bibinfo {author} {\bibfnamefont {S.}~\bibnamefont {Shimoura}},
  \bibinfo {author} {\bibfnamefont {M.}~\bibnamefont {Takechi}}, \bibinfo
  {author} {\bibfnamefont {Y.}~\bibnamefont {Togano}}, \bibinfo {author}
  {\bibfnamefont {H.}~\bibnamefont {Wang}},\ and\ \bibinfo {author}
  {\bibfnamefont {K.}~\bibnamefont {Yoneda}},\ }\bibfield  {title} {\bibinfo
  {title} {{S}hell and shape evolution at ${N}=28$: {T}he $^{40}\mathrm{Mg}$
  ground state},\ }\href {https://doi.org/10.1103/PhysRevC.89.041303}
  {\bibfield  {journal} {\bibinfo  {journal} {Phys. Rev. C}\ }\textbf {\bibinfo
  {volume} {89}},\ \bibinfo {pages} {041303} (\bibinfo {year}
  {2014})}\BibitemShut {NoStop}%
\bibitem [{\citenamefont {Murray}\ \emph {et~al.}(2019)\citenamefont {Murray},
  \citenamefont {MacCormick}, \citenamefont {Bazin}, \citenamefont
  {Doornenbal}, \citenamefont {Aoi}, \citenamefont {Baba}, \citenamefont
  {Crawford}, \citenamefont {Fallon}, \citenamefont {Li}, \citenamefont {Lee},
  \citenamefont {Matsushita}, \citenamefont {Motobayashi}, \citenamefont
  {Otsuka}, \citenamefont {Sakurai}, \citenamefont {Scheit}, \citenamefont
  {Steppenbeck}, \citenamefont {Takeuchi}, \citenamefont {Tostevin},
  \citenamefont {Tsunoda}, \citenamefont {Utsuno}, \citenamefont {Wang},\ and\
  \citenamefont {Yoneda}}]{Mur19a}%
  \BibitemOpen
  \bibfield  {author} {\bibinfo {author} {\bibfnamefont {I.}~\bibnamefont
  {Murray}}, \bibinfo {author} {\bibfnamefont {M.}~\bibnamefont {MacCormick}},
  \bibinfo {author} {\bibfnamefont {D.}~\bibnamefont {Bazin}}, \bibinfo
  {author} {\bibfnamefont {P.}~\bibnamefont {Doornenbal}}, \bibinfo {author}
  {\bibfnamefont {N.}~\bibnamefont {Aoi}}, \bibinfo {author} {\bibfnamefont
  {H.}~\bibnamefont {Baba}}, \bibinfo {author} {\bibfnamefont {H.}~\bibnamefont
  {Crawford}}, \bibinfo {author} {\bibfnamefont {P.}~\bibnamefont {Fallon}},
  \bibinfo {author} {\bibfnamefont {K.}~\bibnamefont {Li}}, \bibinfo {author}
  {\bibfnamefont {J.}~\bibnamefont {Lee}}, \bibinfo {author} {\bibfnamefont
  {M.}~\bibnamefont {Matsushita}}, \bibinfo {author} {\bibfnamefont
  {T.}~\bibnamefont {Motobayashi}}, \bibinfo {author} {\bibfnamefont
  {T.}~\bibnamefont {Otsuka}}, \bibinfo {author} {\bibfnamefont
  {H.}~\bibnamefont {Sakurai}}, \bibinfo {author} {\bibfnamefont
  {H.}~\bibnamefont {Scheit}}, \bibinfo {author} {\bibfnamefont
  {D.}~\bibnamefont {Steppenbeck}}, \bibinfo {author} {\bibfnamefont
  {S.}~\bibnamefont {Takeuchi}}, \bibinfo {author} {\bibfnamefont {J.~A.}\
  \bibnamefont {Tostevin}}, \bibinfo {author} {\bibfnamefont {N.}~\bibnamefont
  {Tsunoda}}, \bibinfo {author} {\bibfnamefont {Y.}~\bibnamefont {Utsuno}},
  \bibinfo {author} {\bibfnamefont {H.}~\bibnamefont {Wang}},\ and\ \bibinfo
  {author} {\bibfnamefont {K.}~\bibnamefont {Yoneda}},\ }\bibfield  {title}
  {\bibinfo {title} {Spectroscopy of strongly deformed $^{32}\mathrm{Ne}$ by
  proton knockout reactions},\ }\href
  {https://doi.org/10.1103/PhysRevC.99.011302} {\bibfield  {journal} {\bibinfo
  {journal} {Phys. Rev. C}\ }\textbf {\bibinfo {volume} {99}},\ \bibinfo
  {pages} {011302} (\bibinfo {year} {2019})}\BibitemShut {NoStop}%
\end{thebibliography}%

\end{document}